\documentclass[aps,prb,amsmath,twocolumn,amssymb,titlepage,superscriptaddress]{revtex4-1}
\usepackage[english]{babel}
\usepackage{graphicx}
\usepackage{transparent}
\usepackage{amsmath}
\usepackage{amssymb}
\usepackage{epstopdf}
\usepackage{amsfonts}
\usepackage{bm}
\usepackage{mathdots}
\usepackage{times}
\usepackage{pdfpages}
\usepackage{mathtools}
\usepackage{stmaryrd}
\usepackage{hyperref}
\usepackage{tabularx}
\usepackage{hhline}

\newcommand{\ket}[1]{|#1\rangle}
\newcommand{\bra}[1]{\langle#1|}

\newcommand{\matele}[3]{\langle #1|#2|#3\rangle}
\newcommand{\I}{\mathrm{i}}

\newcommand{\ave}[1]{\langle #1 \rangle}

\newcommand{\green}[2]{\langle \mathcal{T_C} #1(t) #2(t') \rangle}
\newcommand{\beq}[1]{\begin{equation} #1 \end{equation}}
\newcommand{\bsplit}[1]{\begin{equation} \begin{split} #1 \end{split} \end{equation}}

\newcommand{\astcycl}{\mathrlap{\kern0.085em{\circlearrowright}}\ast}
\newcommand{\taucycl}{\mathrlap{\kern0.42em{\bullet}}\circlearrowright}
\DeclareMathOperator{\Tr}{Tr}

\begin{document}
\title{Dynamics of screening in photo-doped Mott insulators}
\author{Denis Gole\v z}
\affiliation{Department of Physics, University of Fribourg, 1700 Fribourg, Switzerland}
\author{Martin Eckstein}
\affiliation{Max Planck Research Department for Structural Dynamics, University of Hamburg-CFEL, 22761 Hamburg, Germany}
\author{Philipp Werner}
\affiliation{Department of Physics, University of Fribourg, 1700 Fribourg, Switzerland}


\begin{abstract}
We use a nonequilibrium implementation of extended dynamical mean field theory to study the effect of dynamical screening in photo-excited Mott insulators. The insertion of doublons and holes adds low-energy screening modes and leads to a reduction of the Mott gap. The coupling to low-energy bosonic modes furthermore opens new relaxation channels and significantly speeds up the thermalization process. 
We also investigate the effect of the energy distribution of the doped carriers on the screening. 
\end{abstract}
\maketitle

\section{Introduction}
Driving a material out of equilibrium by a strong laser pulse can provide new insights into correlation phenomena,\cite{perfetti2006,dalconte12,fausti2011,gadermaier10,gadermaier2014} and even induce transitions into nonthermal ``dark" states.\cite{stojchevska2014} A conceptually rather simple example of a pulse induced nonequilibrium phase transition is the photo-doping of a Mott insulator.\cite{iwai2003,okamoto2007,wall2011,okamoto2010} In these experiments, a pulse with frequency larger than the Mott gap produces doublon-hole pairs and hence results in a nonthermal conducting state. While the metallization happens on a femto-second timescale, the relaxation back to a Mott insulating state can take several picoseconds.\cite{iwai2003} 

On the theoretical side, different aspects of the photo-doping and thermalization process have been recently investigated. In a system with sufficiently large gap, the relaxation proceeds in two stages. In the first stage, the photo-doped carriers ``thermalize" within the Hubbard band due to electron-electron scattering, or loose their kinetic energy due to scattering with phonons\cite{golez2012a, werner2015} or spins.\cite{golez2014,kogoj2014, eckstein2014} Scattering processes which change the number of doublon-hole pairs, necessary for thermalization, depend exponentially on the gap size.\cite{sensarma2010, eckstein2010} In small-gap insulators, the initial kinetic energy of the photo-doped carriers may be sufficient for ``impact ionization,"\cite{werner2014} which introduces an additional timescale into the problem. 

An important aspect, which has not been considered in these previous studies based on the nonequilibrium dynamical mean field approximation,\cite{aoki2014_rev} is the effect of screening from longer-ranged Coulomb interactions. The interaction parameters used within an (extended) Hubbard-model type description are partially screened interactions, which can be obtained from the fully screened interaction by removing the screening processes withing the subspace of the low-energy model.\cite{aryasetiawan2004} The low-energy screening is however very different in a Mott insulator and in a metal. If mobile carriers are inserted into a Mott insulator by photo-doping, the screening of the Coulomb interaction will change, and this should have a noticeable effect on the nature of the photo-doped state and on the relaxation processes. Other effects of the longer-range Coulomb interaction on the non-equilibrium dynamics have been studied in a one dimensional chain using the time dependent Lanczos method. There, a local enhancement of charge (spin) order\cite{lu2012,lu2013} and an appearance of in-gap states \cite{lu2015} can be observed.

A method which captures the screening from long-range Coulomb interactions is extended dynamical mean-field theory (EDMFT). While this formalism has been developed more than a decade ago,\cite{sengupta1995,si1996,sun2002} an accurate numerical implementation has only recently become feasible.~\cite{werner2007, werner2010, ayral2013} Applications to the Hubbard model with on-site and inter-site interactions have clarified the phase diagram and the dominant low-energy screening modes in the insulating and metallic phase.~\cite{huang2014} Here, we extend EDMFT to the nonequilibrium domain by implementing the scheme on a Kadanoff-Baym contour. This allows us to study the dynamical screening effect after a photo-doping excitation in real time. We show that the screening, and the associated possibility to emit and absorb plasmons, influences the thermalization process in significant ways and that the doping-induced screening of the Coulomb interaction has an effect on the gap size of photo-doped Mott insulators. We also show evidence for nontrivial transient states induced by time-dependent changes of the screening environment.

This paper is organized as follows: In section II we discuss the nonequilibrium generalization of EDMFT and its implementation using an impurity solver which combines the non-crossing approximation with a weak-coupling expansion for the retarded interaction. In section III we analyze the relaxation dynamics of doublons after a photo-doping pulse, the doping-induced changes in the electron spectral function and in the screening modes, and the effect of the energy distribution of the carriers on the screening. While most of the results pertain to Mott insulators, we also consider the photo-doping of strongly correlated metals and in particular the effect of the destruction of the quasi-particle peak on the screening. Section IV is a conclusion and outlook. 
\section{Model and method}

\subsection{$U$-$V$ Hubbard model}

We consider the single-band $U$-$V$ Hubbard model on the two-dimensional square lattice
\begin{eqnarray}
  &&H(t)=-\sum_{\langle ij \rangle\sigma}s_{ij}(t)(c_{i\sigma}^{\dagger}c_{j\sigma}+h.c.)-\mu \sum_{i}n_i  \nonumber\\
  &&+\sum_{i} U(t) (n_{i\uparrow}-\tfrac{1}{2})(n_{i\downarrow}-\tfrac{1}{2})\nonumber + \sum_{\langle ij\rangle} V(t)(n_{i}-1)(n_{j}-1),
  \label{Eq.:Hubbard-UV}
\end{eqnarray}
where $c_{i\sigma}$ denotes the annihilation operators of a fermion with spin $\sigma$ at the lattice site $i$, $n_{i}=n_{i\uparrow}+n_{i\downarrow}$, $s(t)$ is the hopping integral between neighbouring sites, whose time-dependence captures the effect of an in-plane electric field, $\mu$ is the chemical potential, $U$ the on-site interaction energy, and $V$ the interaction energy between two electrons on neighbouring sites. The case  $V=0$ corresponds to the conventional Hubbard model. Using the identity $(n_i-1) (n_i-1)=2 (n_{i\downarrow}-1/2) (n_{i\uparrow}-1/2)+1/2$ we can combine the interaction terms  as $\frac{1}{2}\sum_{ij}v_{ij}\bar n_{i}\bar n_{j}$  with $\bar n=n-1$  the density fluctuation operator (for the case of half-filling),
\bsplit{
    v_{ij}=U\delta_{ij}+V\delta_{\langle ij\rangle},
}
and a shift of the chemical potential $\mu\rightarrow \tilde \mu=\mu+U/2$.

The grand-canonical partition function is $\mathcal{Z}=\text{Tr}[\mathcal{T_C} e^{S}]$, where $\mathcal{T_C}$ denotes the countour-ordering operator for the Kadanoff-Baym contour which runs from time $0$ to some maxium simulation time $t_\text{max}$ along the real time axis, back to $0$ along the real-time axis, and then to $-i\beta$ along the imaginary-time axis. 
Following Ref.~\onlinecite{ayral2013}, we express $\mathcal{Z}$ as a coherent-state path integral, $\mathcal{Z}=\int \mathcal{D}[c^{*}_{i}, c_i]e^{S}$, with the action given by
\begin{align}
  S[c^*,c]=&-\I  \Bigg\{ \int_\mathcal{C} dt \sum_{ij\sigma} c_{i\sigma}^{*}(t) [(-\I\partial_{t}-\tilde \mu)\delta_{ij}+s_{ij}(t)] c_{j\sigma}(t)  \nonumber\\
  & +\frac{1}{2} \sum_{ij} \bar n_{i}(t) v_{ij} \bar n_{j}(t) \Bigg\}.
\end{align}
In order to decouple the interaction term we will use the Hubbard-Stratonovich identity
\bsplit{
  & \exp\left(\I\frac{1}{2} \int_\mathcal{C} dt dt'  \sum_{ij} b_{i}(t)A_{ij}(t,t') b_{j}(t') \right)= \\
  & \int \frac{\mathcal{D}[x_1(t),x_2(t), \ldots]}{\sqrt{(2\pi)^N \text{det} A}}\\
   & \times \exp \Bigg(\I \Bigg[ \int_\mathcal{C} dt dt' \Big\{ -\frac{1}{2}\sum_{ij}x_{i}(t) [A^{-1}]_{ij}(t,t') x_{j}(t')  \\
   &  - \sum_i x_{i}(t) b_{i}(t) \delta_\mathcal{C}(t,t') \Big\} \Bigg]\Bigg),
}
where $A$ is a real symmetric positive-definite matrix and $b_{i}(t)$ and $x_{i}(t)$ are fields defined on the contour. We will perform the so-called ``$UV$-decoupling",\cite{ayral2013} where the full interaction term is decoupled via an auxiliary bosonic field $\phi_i$.  Choosing $b_i=\I \bar n_i$, $A_{ij}=v_{ij}$ and $x_i=\phi_i$ the transformed action becomes
\bsplit{
  S[c^*,c,\phi]=- \I \int_\mathcal{C} dt dt' \Big[ -\sum_{ij\sigma}c_{i\sigma}^{*}(t) [(G_0^H)^{-1}]_{ij}(t,t') c_{j\sigma}(t')  \\
  + \frac{1}{2}\sum_{ij} \phi_i(t) [v^{-1}]_{ij}\delta_\mathcal{C}(t,t') \phi_j(t') +\I \sum_{i} \phi_{i}(t)  \delta_\mathcal{C}(t,t')\bar n_{i}(t) \Big],
  \label{Eq.:Action_lattice}
}
where we have introduced the fermionic Green's function $[(G_{0}^H)^{-1}]_{ij}=[(\I\partial_{t}+\tilde \mu)\delta_{ij}-s_{ij}]\delta_c(t,t')$. The fermionic and bosonic Green's functions for this action are
\bsplit{
  & G_{ij}(t,t')=-\I \ave{c_i(t) c_j^{*}(t')} \\
  & W_{ij}(t,t')=\I \ave{\phi_i(t) \phi_j(t')}, \\
}
with the expectation values defined as $\langle \ldots \rangle =\frac{1}{\mathcal{Z}}\int \mathcal{D}[c^{*}_{i}, c_i] [e^{S} \ldots]$.

\subsection{Nonequilibrium EDMFT}
 The EDMFT approximation maps the lattice problem onto a single-site effective action 
\bsplit{
 S_\text{imp}^{e-b}[c^*,c,\phi]&=-\I \int_\mathcal{C} dt dt' \Big\{ -\sum_{\sigma} c_{\sigma}^{*}(t) \mathcal{G}_{0\sigma}^{-1}(t,t') c_{\sigma}(t') \\
 & + \frac{1}{2}\phi(t)\mathcal{U}^{-1}(t,t')\phi(t') +\I \phi(t)\delta_\mathcal{C}(t,t')\bar n(t')\Big\},
\label{Eq.:Action_EDMFTwBoson}
}
whose fermionic ($\mathcal{G}_0(t,t')$) and bosonic ($\mathcal{U}(t,t')$) Weiss fields are fixed by a self-consistency condition.
This effective action 
is obtained by integrating out all sites but one from the lattice action (\ref{Eq.:Action_lattice}) and taking the infinite dimensional limit.\cite{georges1996,ayral2013} The hybridization function for the electrons, $\Delta_\sigma(t,t')$, is given  by
  $ \mathcal{G}_{0\sigma}^{-1}(t,t')=[\I \partial_t+\tilde \mu]\delta_\mathcal{C}(t,t')-\Delta_\sigma(t,t')$
and the equivalent bosonic function $\mathcal{D}$ corresponds to the retarded component of the interaction $\mathcal{U}$: $\mathcal{U}(t,t')=U(t)\delta_\mathcal{C}(t,t')+\mathcal{D}(t,t')$. In order to obtain a purely electronic action we can integrate out the bosonic field $\phi$ and obtain:
\beq{
  \int \mathcal{D}[\phi] e^{S_{imp}^{e-b}}=e^{S^e} e^{\frac{1}{2}\text{Tr}[\ln\mathcal{U}]},\label{Z_e}
}
where electronic action $S_e$ is given by
\bsplit{
 S_\text{imp}^{e}[c^*,c]=&-\I \int_\mathcal{C} dt dt' \Big\{ -\sum_{\sigma}c_{\sigma}^{*}(t) \mathcal{G}_{0\sigma}^{-1}(t,t') c_{\sigma}(t') \\
 &+ \frac{1}{2} \bar n(t)\mathcal{U}(t,t') \bar n(t') \Big\} .
\label{Eq.:Action_EDMFT}
}

In EDMFT, the impurity Dyson equations for $G_\text{imp}(t,t')=-\I \langle {c}(t){c^{*}(t')}\rangle_{S_\text{imp}^{e-b}}$ and $W_\text{imp}=\I \langle{\phi}(t){\phi(t')}\rangle_{S_\text{imp}^{e-b}}$ are given by 
\begin{align}
  G_\text{imp}&=\mathcal{G}_0+\mathcal{G}_0*\Sigma*G_\text{imp}, \label{Eq.:Dyson_impurity_G}\\
  W_\text{imp}&=\mathcal{U}+\mathcal{U}*\Pi*W_\text{imp}, \label{Eq.:Dyson_impurity_W}
\end{align}
where $\Sigma$ $(\Pi)$ is the fermionic (bosonic) self-energy and the star denotes the convolution on the contour. The bosonic propagator $W_\text{imp}$ and the retarded interaction $\mathcal{U}$ are connected through the charge-charge correlator $\chi_\text{imp}(t,t')=\green{\bar n}{\bar n}$ as (see App.~\ref{App.:Boson_from_chi})
\beq{
  W_\text{imp}=\mathcal{U}-\mathcal{U}*\chi_\text{imp}*\mathcal{U}.
  \label{Eq.:Bosonic_from_cc}
}

The solution of the impurity problem, i.e., the calculation of $G_\text{imp}$ and $W_\text{imp}$ is described in Sec.~\ref{Sec.:Impurity}. From the Dyson equations (\ref{Eq.:Dyson_impurity_G}) and (\ref{Eq.:Dyson_impurity_W}) we obtain the self-energies $\Sigma$, $\Pi$ and the EDMFT approximation identifies these with the lattice self-energies. The solution of the lattice Dyson equations
\begin{align}
  G_k&=G_{0,k}+G_{0,k}*\Sigma* G_k, \label{Eq.:Dyson_G}\\
  W_k&=v_k+v_k*\Pi* W_k ,\label{Eq.:Dyson_W}
\end{align}
then yields an approximation for the lattice Green's functions $G_{k},W_k$ and from these we can estimate the local lattice Green's functions $G_\text{loc}$ and $W_\text{loc}$ by averaging over $k$. The EDMFT self-consistency condition demands that $G_\text{loc}=G_\text{imp}$ and $W_\text{loc}=W_\text{imp}$. Therefore, updated Weiss fields can be obtained from the impurity Dyson equations (\ref{Eq.:Dyson_impurity_G}) and (\ref{Eq.:Dyson_impurity_W}) by replacing the impurity Green's functions with the local lattice Green's functions. The solution of the impurity problem (\ref{Eq.:Action_EDMFT}) via (\ref{Eq.:Bosonic_from_cc}), the impurity and lattice Dyson equations (\ref{Eq.:Dyson_impurity_G}), (\ref{Eq.:Dyson_impurity_W}) and (\ref{Eq.:Dyson_G}), (\ref{Eq.:Dyson_W}), the EDMFT approximation for the lattice self-energies and the EDMFT self-consistency equations for $G_\text{loc}$ and $W_\text{loc}$ 
form 
the closed set of nonequilibrium EDMFT equations.

The nonequilibrium EDMFT calculation is implemented as a step-wise time-propagation, which starts from an equilibrium EDMFT solution at time $t=0$. For each time-step along the real-time axis, we iterate the following procedure until convergence is reached:
\begin{enumerate}
\item Start with some initial guess for the 
dynamical mean fields $\Delta(t,t')$ and $\mathcal{D}(t,t')$ (for example extrapolations of the converged solution for the previous time step),
\item Solve the impurity problem and obtain $G_\text{imp}(t,t')$ and $W_\text{imp}(t,t')$ as described in the Sec.~\ref{Sec.:Impurity}, 
\item Obtain a new approximation for the 
dynamical mean fields 
by closing the lattice self-consistency relations as described in the Sec.~\ref{Sec.:Lattice}.
\end{enumerate}

Since we will use a strong-coupling impurity solver,\cite{eckstein2010} the implementation of the self-consistency loop needs to be slightly reformulated. While the scheme for the fermionic self-consistency loop has been discussed in Ref.~\onlinecite{aoki2014_rev}, we will explain the implementation of the bosonic self-consistency loop in Sec.~\ref{Sec.:Lattice}.

\subsection{Impurity solver}\label{Sec.:Impurity}

In order to solve the impurity problem corresponding to the action (\ref{Eq.:Action_EDMFT}) we combine a hybridization expansion and a weak-coupling expansion in powers of the retarded density-density interaction.
It is therefore convenient to first express the partition function in terms of $\Delta$ and $\mathcal{D}$ as $Z=\text{Tr}_c[\mathcal{T}_\mathcal{C}e^\mathcal{S}]$, with the contour action
\begin{align}
\mathcal{S}=&-\I\Bigg\{\int_\mathcal{C} dt dt' \sum_\sigma c^\dagger_\sigma(t)\Delta_\sigma(t,t')c_\sigma(t')\nonumber\\
&+\frac{1}{2}\int_\mathcal{C}dt dt' \bar n(t)\mathcal{D}(t,t')\bar n(t')
+\int_\mathcal{C} dt H_\text{loc}(t)+\text{const.}\Bigg\}
\end{align}
and $H_\text{loc}(t)=-\tilde \mu \sum_\sigma \bar n_\sigma(t)+U(t)\bar n_\uparrow(t) \bar n_\downarrow(t).$ 
The double expansion then leads to 
\begin{align}
  &Z  = \sum_{n=0}^{\infty} \sum_{m=0}^{\infty} \frac{(-\I)^n}{n!} \frac{(-\I)^m}{m!}\sum_{\sigma_1\ldots\sigma_n} \text{Tr} \Bigg[ \nonumber\\
  &\times \int dt_1 \ldots  dt_{n'} \int d\tilde t_1 \ldots d\tilde t_{m'} T_\mathcal{C} e^{-\I\int_\mathcal{C} dt H_\text{loc}(t)} \nonumber\\
  &\times c^\dagger_{\sigma_1}(t_1) c_{\sigma_1}(t'_{1}) \ldots c^\dagger_{\sigma_n}(t_{n}) c_{\sigma_n}(t'_{n})\nonumber\\
  &\times \bar n(\tilde t_{1}) \bar n(\tilde t'_{1}) \ldots \bar n(\tilde t_{m})\bar n(\tilde t'_{m})\nonumber\\
  &\times \Delta_{\sigma_1}(t_1,t_1')\ldots \Delta_{\sigma_n}(t_n,t_n')     \mathcal{D}(\tilde t_1,\tilde t_1')\ldots \mathcal{D}(\tilde t_m,\tilde t_m')
  \Bigg].
  \label{Eq.:Action_expansion}
\end{align}
  
\begin{figure}[t]
\includegraphics[scale=0.3]{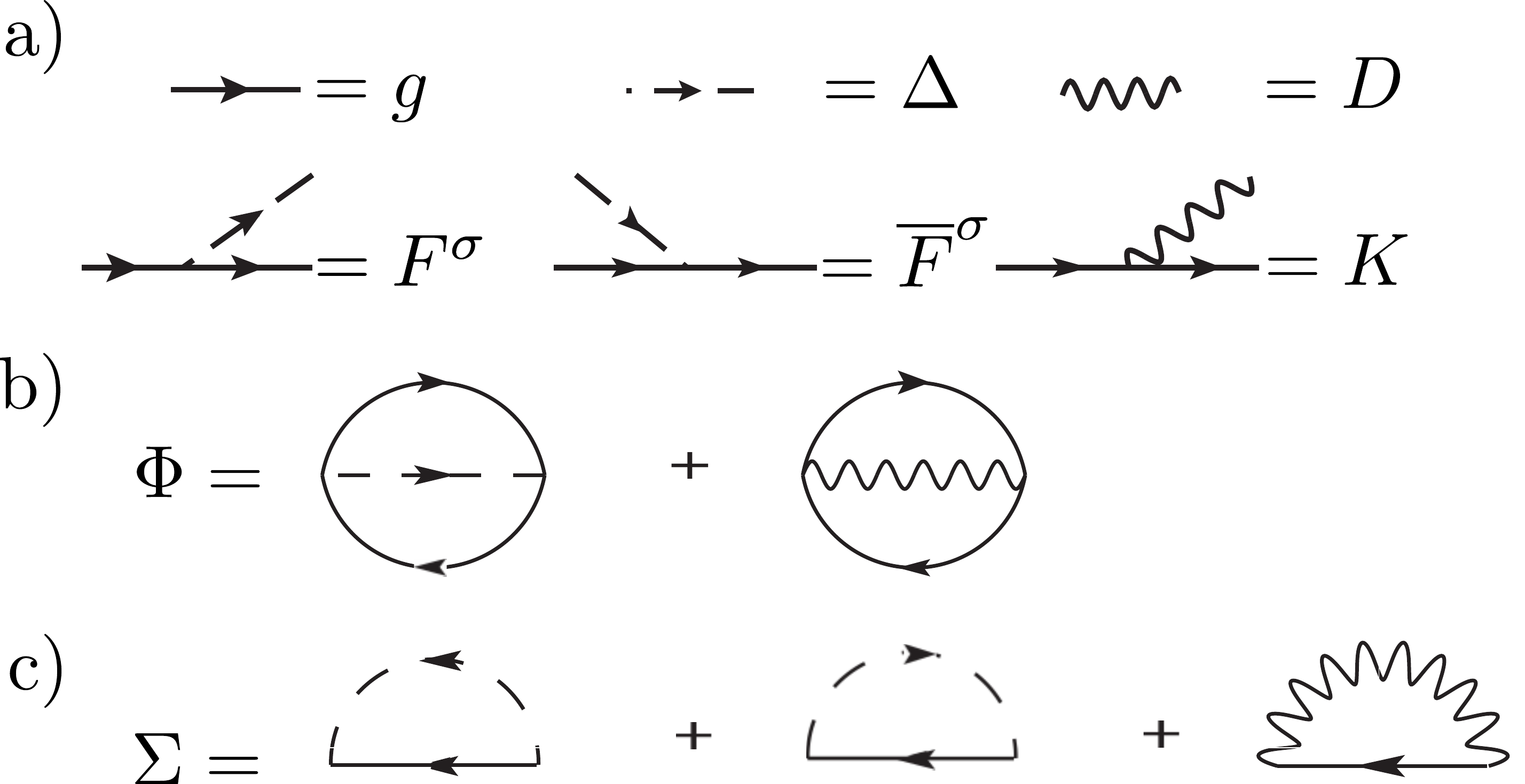}\\
\caption{ 
(a) Vertices representing the coupling between pseudo-particle propagators (solid lines) and hybridization functions (dashed lines), or between pseudo-particle propagators and retarded interactions (wavy lines). 
(b) 
The lowest order diagrams (in $\Delta$ and $\mathcal{D}$) contributing to the Luttinger-Ward functional.
(c) The pseudo-particle self-energy diagrams corresponding to the approximation (b) for the Luttinger-Ward functional. }
\label{Fig:Diagrams}
\end{figure}

In order to evaluate the trace over the electronic configurations one can insert a complete set of eigenstates of $H_\text{loc}$,  $\sum_n \ket{n}\bra{n}$, between consecutive operators $O$ and factor the trace into a product of impurity propagators $g$ and hybridization vertices for electrons ($F^{\sigma}$)  or bosons ($K$):
\bsplit{
  g_{n}(t,t')&=-\I\matele{n}{\mathcal{T}_c e^{-\I\int_{t'}^{t} d\bar t H_\text{loc}(\bar t) }}{n}, \\
  F_{nm}^{\sigma}&=\matele{n}{c_{\sigma}}{m}, \\
  K_{nm}&= \delta_{nm} \matele{n}{\bar n}{n}.
  }
The Taylor expansion of the partition function can then be represented as the sum of all diagrams made up of  bare impurity propagators and vertices connected by $\Delta$ and $\mathcal{D}$ lines, in analogy to the discussion for the Hubbard model in Ref.~\onlinecite{aoki2014_rev}. The vertices corresponding to the coupling between the  pseudo-particles and the hybridization function $\Delta$, and the coupling between pseudo-particles and the retarded interaction $\mathcal{D}$ are shown Fig.~\ref{Fig:Diagrams}(a).

To resum the series we define the pseudo-particle self-energy $\Sigma_p$ as a sum of all parts of the above diagrams, that cannot be separated into two by cutting one pseudoparticle propagator line. With this we can define the renormalized pseudoparticle propagator $\mathcal{G}$ via the pseudo-particle Dyson equation
  \beq{
    \mathcal{G}=g+g \astcycl \Sigma_p \astcycl \mathcal{G},
  }
 where 
 the star with arrow 
 denotes the cyclic convolution $\int_\mathcal{C} d\bar t a(t,\bar t) b(\bar t ,t')$ restricted to countour-ordered time arguments $t>\bar t>t'$ (cyclic time ordering), for details see Ref. \onlinecite{eckstein2010}. 

In practice, we need to truncate the self-energy expansion at a given order in $\Delta$ and $\mathcal{D}$. 
To obtain a conserving approximation, we construct a Luttinger-Ward functional $\Phi$ from all vacuum skeleton diagrams involving fully renormalized pseudoparticles propagators and the corresponding vertices. The lowest order perturbative strong-coupling method is called noncrossing approximation (NCA),\cite{grewe1981,coleman1984} because it sums all diagrams without crossing $\Delta$ and/or $\mathcal{D}$ lines. We will use this approximation in the present study for the nonequilibrium case.

A detailed derivation of the strong-coupling equations on the Keldysh countour has been presented for the Hubbard model in Refs.~\onlinecite{eckstein2010,aoki2014_rev}. In this case, the NCA diagrams for $\Phi$ contain just one $\Delta$ line.  The generalisation to the partial resummation of the series given in Eq.~(\ref{Eq.:Action_expansion}) includes 
an additional diagram with a single $\mathcal{D}$ line, see 
Fig.~\ref{Fig:Diagrams}(b). The expansion of the partition function in Eq.~(\ref{Eq.:Action_expansion}) includes terms with crossing electron and boson propagators, but the lowest order $\Phi$ diagram which produces these terms is $O(K^4 F^2)$ or $O(F^2 K^4)$ and will be neglected in our calculations. 

The self-energies are obtained as a functional derivative of the Luttinger-Ward functional with respect to the corresponding pseudoparticle propagators, 
\beq{
  \Sigma_p(t,t')=\frac{\delta \Phi}{\delta \mathcal{G}(t',t)}, 
}
and are depicted for the NCA case in Fig.~\ref{Fig:Diagrams}(c). 
The explicit expression for these diagrams is
\bsplit{
  \Sigma_p(t,t') =& -\I[F^{\sigma} g(t,t')\bar F^{\sigma} \Delta_{\sigma}(t,t')+F^{\sigma} g(t,t')\bar F^{\sigma} \Delta_{\sigma}(t',t)]\\
  &-\I[K g(t,t') K D(t',t)].
\label{Eq.:Sigma_NCA}
}

\subsection{Lattice Dyson equation}\label{Sec.:Lattice}

We need to solve the lattice Dyson equations in order to obtain new approximations for the Weiss functions. In EDMFT, the impurity and the lattice Dyson equations for the fermions read
\bsplit{
 &  G_\text{imp}^{-1}(t,t')=\mathcal{G}_0^{-1}(t,t')-\Sigma(t,t'), \\
 & G_{k}^{-1}(t,t')=(\I\partial_t +\tilde \mu)\delta_\mathcal{C}(t,t')-E_k(t,t')-\Sigma(t,t'),
  \label{Eq.:Lattice_Dyson_eq_fermion}
}
where we have introduced $E_k(t,t')=\epsilon_k\delta(t,t')$. In order to close the self-consistency loop for the electrons we can proceed as described in Sec.~II.B.4 of Ref.~\onlinecite{aoki2014_rev}, but the bosonic self-consistency loop requires some modifications.

The impurity and the lattice Dyson equations for the bosons read
\bsplit{
    W_\text{imp}^{-1}(t,t')= \mathcal{ U}^{-1}(t,t')-\Pi(t,t'),\\
    W_{k}^{-1}(t,t')= v_k^{-1}(t,t')-\Pi(t,t'),
}
where the Weiss field is given by $\mathcal{U}(t,t')=\mathcal{U}_0(t,t')+\mathcal{D}(t,t') $ and $\mathcal{U}_0(t,t')=U\delta_\mathcal{C}(t,t').$ Note that the inverse of the bosonic Weiss field appears in the Dyson equation, so we have to 
calculate 
$\mathcal{U}^{-1}.$ We make the ansatz $\mathcal{  U}^{-1}= \mathcal{U}_0^{-1}-B$. Since $\mathcal{ U}*\mathcal{ U}^{-1}=\delta_\mathcal{C}$, the components $ \mathcal{U}_0^{-1}$ and $B$ have to satisfy the condition
$(\mathcal{U}_0+\mathcal{D})*(\mathcal{U}_0^{-1}-B)=\delta_\mathcal{C}$. From this and the relation  $\mathcal{U}_0^{-1}=\frac{1}{U}\delta_\mathcal{C}(t,t')$ one finds 
\bsplit{
& \left[\delta_\mathcal{C}(t,\bar t) +  \frac{1}{U(t)} \mathcal{D}(t,\bar t)\right]*B(\bar t,t')=\frac{1}{U(t)} \mathcal{D}(t,t') \frac{1}{U(t')},
\label{Eq.:Inverse_bosonic}
}
so that the equation for $B(t,t')$ is a numerically stable integral equation. 

Next, we 
define the bosonic function 
$u^{-1}(t,t')=\mathcal{U}_0^{-1}(t,t')-\Pi(t,t')$, which allows us to write the impurity Dyson equation as 
\begin{align}
  &  W_\text{imp}^{-1}(t,t') 
  = u^{-1}(t,t')-B(t,t'), \\
  &  (1+W_\text{imp}*B)* u= W_\text{imp}. \label{Eq.:Impurity_Dyson_bosons}
\end{align}
In the second line, we thereby obtained a numerically stable integral equation for $u$. In solving this integral equation, instantaneous terms proportional to $\delta_\mathcal{C}(t,t')$  need to be treated separately; such terms arise because $W_\text{imp}(t,t')=U(t)\delta_\mathcal{C}(t,t')+W^\text{reg}(t,t')$ has an instantaneous term. In the general case, where the solution may contain a singular term we write: $[(1+F^{\delta}(t))\delta(t,\bar t)+F^\text{reg}(t,\bar t)]*[G^{\delta}\delta_\mathcal{C}(\bar t,t')+G^\text{reg}(\bar t,t')]=Q^\delta(t) \delta_\mathcal{C}(t,t')+Q^\text{reg}(t,t')$, where $X^{\delta}$ $(X^\text{reg})$ marks the instantaneous (regular) part of the propagator $X=F,G,Q$. Collecting the instantaneous and retarded terms yields the two equations
\begin{align}
  &G^{\delta}(t)=[1+F^{\delta}(t)]^{-1}Q^{\delta}(t), \label{eq_Gdelta}\\
  &\Bigg[1+[1+F^{\delta}(t)]^{-1}F^\text{reg}(t,\bar t)\Bigg]*G^\text{reg}(\bar t,t')=\nonumber\\
  &\hspace{5mm} [1+F^{\delta}(t)]^{-1} Q^\text{reg}(t,t')-[1+F^{\delta}(t)]^{-1} F^\text{reg}(t,t')G^{\delta}(t'). \label{eq_Greg}
\end{align}
In the case of Eq.~(\ref{Eq.:Impurity_Dyson_bosons}) the propagators $G$, $Q$ and $F$ are
  $G= u$, 
  $Q= W_\text{imp}$, and 
  $F(t,\bar t)=U(t)B(t,\bar t) +  W^\text{reg}(t,t_1)*B(t_1,\bar t)$,  
which means that $F$ has no instantaneous contribution ($F^\delta=0$). The solution is thus obtained from
\begin{align}
  & u^{\delta}(t)=W^{\delta}(t)=U(t), \\
  & [1+F]*u^\text{reg}=  W_\text{imp}^\text{reg}(t,t')-F(t,t')U(t').
\end{align}

\begin{figure*}[ht]
\includegraphics{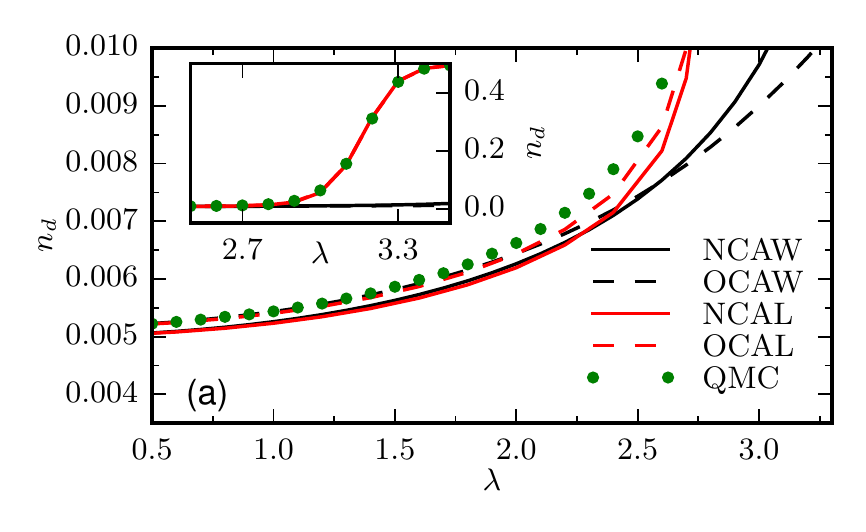}
\includegraphics{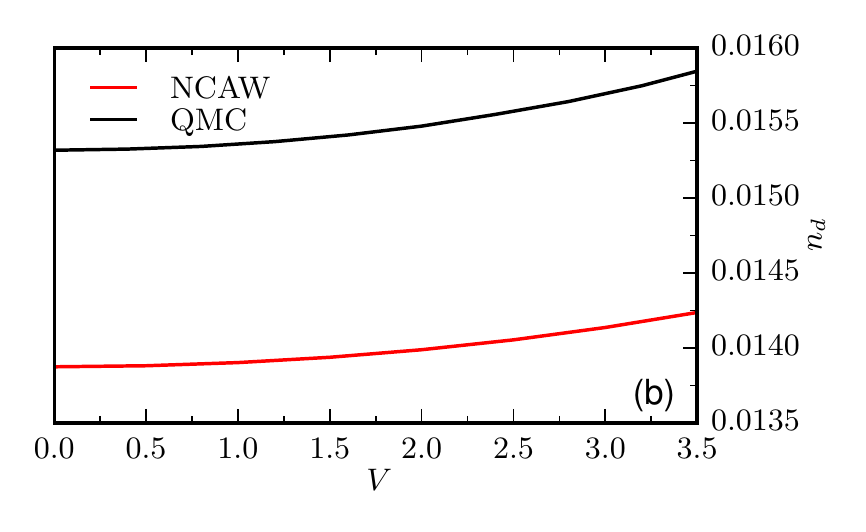} \\
\caption{ 
(a) Comparison of the double occupancy in the Holstein-Hubbard model on the Bethe lattice for $\beta=5$, $U=10$, and 
$\omega_0=2$ obtained from the NCA and OCA approximation in combination with the weak-coupling expansion (W) and Lang-Firsov (L) transformation, and the numerically exact QMC results.
(b) Comparison of the double occupancy for the extended Hubbard model for $\beta=5$, $U=12$ obtained from the NCA approximation and the numerically exact QMC solver on the square lattice.}
\label{Fig:Eq_comparison}
\end{figure*}

The lattice Dyson equation can be rewritten as
  $W_k^{-1}(t,t')= v_k^{-1}(t)\delta_\mathcal{C}(t,t')-\Pi(t,t')= \frac{1}{U+v_k^\text{nonloc}}(t)\delta_\mathcal{C}(t,t') -\Pi(t,t') = u^{-1}(t,t')-A_k(t,t')$,
with $ A_k(t,t')=a_k(t)\delta_\mathcal{C}(t,t')$ and $a_k(t)=\frac{v_k^\text{nonloc}}{U(U+v_k^\text{nonloc})}(t)$ This allows us to cast the problem into the form of a stable Volterra integral equation for $ W_k(t,t'):$
\begin{align}
  (1- u* A_k)* W_k= u.  \label{Eq.:Lattice_Dyson_eq_bosons}
\end{align}
We now use again Eqs.~(\ref{eq_Gdelta}) and (\ref{eq_Greg}) with the substitutions $G= W_k$, $Q= u$ and $F=- u *  A_k=-U(t) a_k(t)\delta_c(t,\bar t)-u^\text{reg}(t,\bar t)a_k(\bar t)$:
\bsplit{
  & W_k^{\delta}(t)=\frac{U(t)}{1-U(t)a_k(t)}=(U+v^\text{nonloc}_k)(t), \\
  & \Bigg[1-\frac{u^\text{reg}(t,\bar t)a_k(\bar t)}{1-U(t) a_k(t)}\Bigg]*W_k^\text{reg}=\frac{u^\text{reg}(t,t')}{1-U(t) a_k(t)}+ \\
  & \hspace{30mm}\frac{u^\text{reg}(t,t')a_k(t')(U+v_k)(t')}{1-U(t)a_k(t)}.
}
The sum over $k$ for the instantaneous term $W_k^\delta$ gives the correct instantaneous contribution for the local bosonic propagator: $\sum_k (U+v^\text{nonloc}_k) = U.$  In analogy to the electronic case we take the sum over $k$ in Eq.~(\ref{Eq.:Lattice_Dyson_eq_bosons}) and use Eq.~(\ref{Eq.:Impurity_Dyson_bosons}) to obtain $W_1$:
\begin{equation}
  B*W_\text{imp}=\sum_k  A_k*W_k \equiv W_1, \label{Eq.:Lattice_Dyson_aux}
\end{equation}
where in the middle expression the instantaneous contribution vanishes, $\sum_k A_k (U+v^\text{nonloc}_k)=\sum_k v^\text{nonloc}_k/U=0$, in agrement with the left hand side. After inserting the conjugate of Eq.~(\ref{Eq.:Lattice_Dyson_eq_bosons}), namely $W_k=u+W_k*A_k*u$, and $W_\text{imp}=u+W_\text{imp}*B*u$ into Eq.~(\ref{Eq.:Lattice_Dyson_aux}) we find
\beq{
  B+B*W_\text{imp}*B=\sum_{k} [A_k+ A_k*W_k*A_k]=W_2,
}
where the instantaneous contribution to the middle expression vanishes due to 
$  \sum_k \frac{v_k }{(U+v_k)U}+\sum_k {v_k^2}{U(U+v_k)}=\sum_k \frac{v_k}{U}=0$. 
The regular part $B$ of the new $\mathcal{U}^{-1}$ can now be calculated from
\beq{
  [1+W_1]*B=W_2.
} 
We still need to obtain the expression for the retarded interaction $\mathcal{D}(t,t')$ and therefore we once more use Eq.~(\ref{Eq.:Inverse_bosonic}) in the form 
\beq{
  (1-U(t) B(t,\bar t) )*\mathcal{D}(\bar t,t')=U(t) B(t,t') U(t'),
}
which is again a stable Volterra integral equation.

\begin{figure*}[ht]
\includegraphics{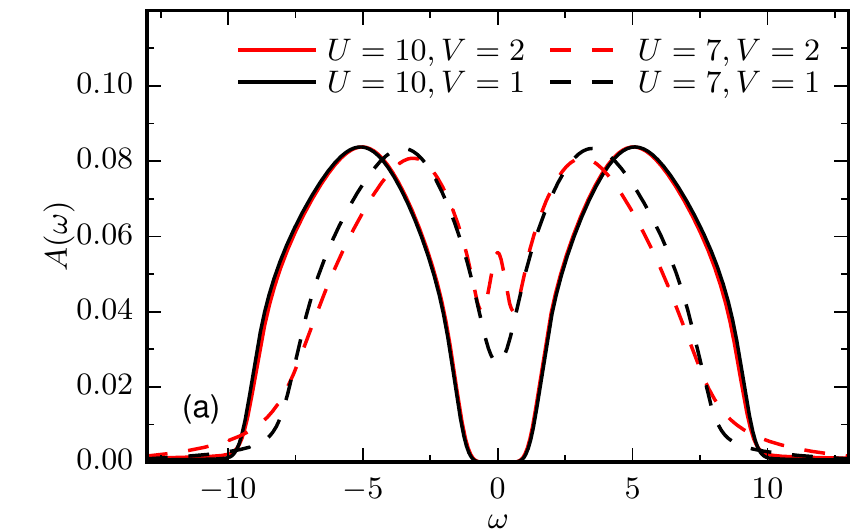}
\includegraphics{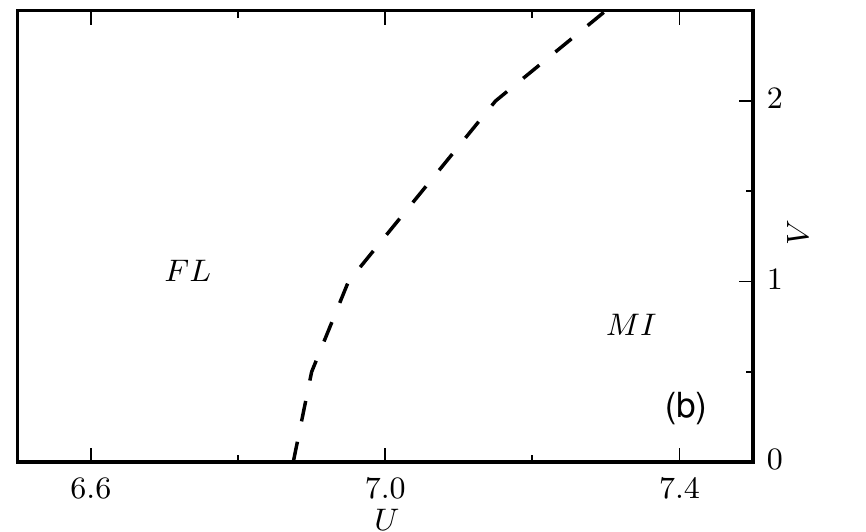}\\
\includegraphics{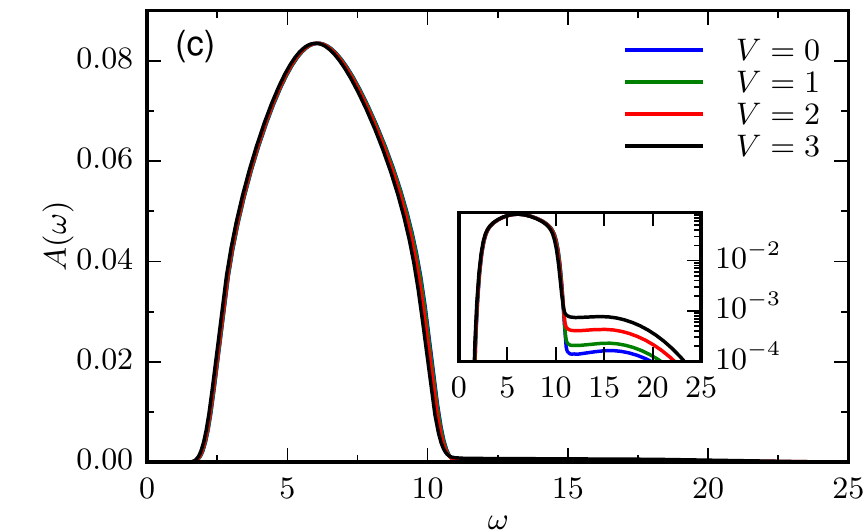}
\includegraphics{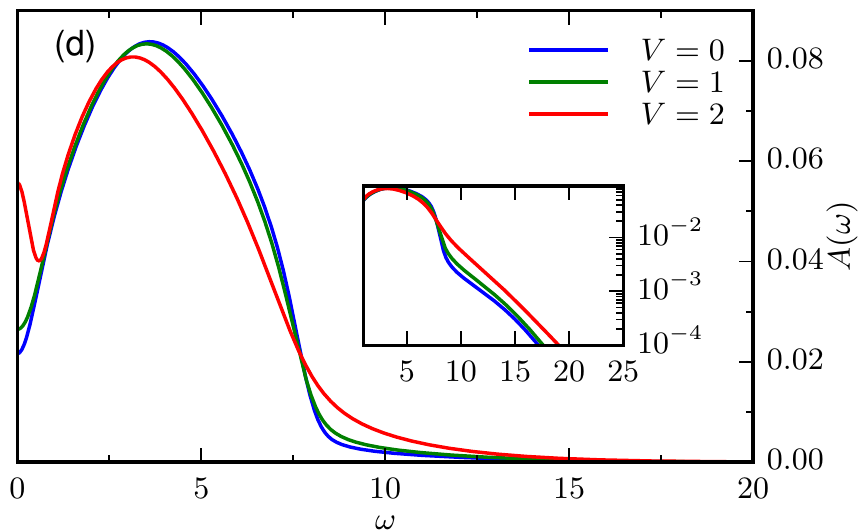}
\caption{
(a) Equilibrium spectral functions for $\beta=20$ and indicated values of $U$ and $V$. (b) Equilibrium phase diagram in the space of $U$ and $V$ at $\beta=20$. The dashed line roughly indicates the metal-insulator crossover defined by the appearance of a quasi-particle peak.
(c)-(d) The equilibrium spectral function $A(\omega)$ for $U=12$ (c) and $U=7$ (d) for indicated values of the nearest neighbor interaction $V$. The insets show the spectra on a logarithmic scale.}
\label{Fig:phase}
\end{figure*}

\subsection{Benchmarks in equilibrium}
To test the accuracy of the impurity solver, we compare some equilibrium results for the double occupancy with numerically exact Monte Carlo data.\cite{werner2010} 
Instead of an EDMFT solution for the extended Hubbard model, we first consider DMFT results for the Holstein-Hubbard model. 
In this case,  there is only one bosonic mode with frequency $\omega_0$ and coupling strength $\lambda$. The bosonic Weiss field is given by the single bosonic mode propagator $\mathcal{D}(t,t')= \lambda^2 \mathcal{D}_0(t,t')=-\I \lambda^2 \Tr[e^{-\I \int_\mathcal{C} dt 
\omega_0 (\phi^2+\Pi^2)/2 } \phi(t) \phi(t') ]$, 
where $\Pi$ denotes the conjugate momentum. 
The black solid and dashed lines in panel (a) of Fig.~\ref{Fig:Eq_comparison} show the double occupation for $\beta=5$, $U=10$, and $\omega_0=2$ obtained from the combined hybridization and weak-coupling expansion, with the solid line corresponding to the NCA approximation and the dashed curve to the one-crossing approximation (OCA), which considers self-energy diagrams with at most one crossing $\Delta$ and/or $\mathcal{D}$ line. The Monte Carlo result, which considers all relevant diagrams, is shown by the green dots. 

For $U=10$, the model without phonon coupling is in the Mott insulating phase. As $\lambda$ is increased, the Coulomb interaction gets screened and the double occupancy increases. Around $\lambda\approx3.0$, the solution crosses over to the bipolaronic insulating phase, which is marked by a large value of the double occupancy, see inset of Fig.~\ref{Fig:Eq_comparison}(a). (At lower temperature, the transition to the bipolaronic insulator occurs via an intermediate metallic phase, see Ref.~\onlinecite{werner2010}.) As expected, the NCA approximation overestimates the correlation effects and thus underestimates the double occupation, while the OCA approximation is quantitatively more accurate. We also see that the weak-coupling treatment of the electron-phonon interaction correctly captures the screening effect, i.e. while the double occupancy is shifted to lower values (as expected in an NCA / OCA calculation), it exhibits the correct $\lambda$-dependence in the weak-coupling regime. The method however does not capture the transition into the bipolaronic state.

For comparison, we also show with red lines the equilibrium results from an alternative scheme in which the phonons are first decoupled by a Lang-Firsov transformation and then integrated out. The Monte Carlo method is actually based on such a decoupling,\cite{werner2007, werner2010} and the NCA/OCA approximation of this Lang-Firsov approach has been discussed in Ref.~\onlinecite{werner2013}. In the small-$\lambda$ regime, this approximate method is slighlty less accurate than the combined hybridization and weak-coupling expansion discussed in this work. On the other hand, the Lang-Firsov method is not limited to small phonon couplings and correctly captures the crossover to the bipolaronic phase. 
 
We next test the EDMFT solution for the $U$-$V$ Hubbard model.
The comparison of the double occupancy $n_d$ to the QMC results shows a similar trend as found in the Holstein-Hubbard calculations, see Fig.~\ref{Fig:Eq_comparison}(b). The NCA approximation underestimates the double occupation, but correctly captures the screening effects in the weak coupling regime $V\lesssim 3.5$. 

Since the combined NCA and weak-coupling approach provides a qualitatively correct description of the physics in the Mott insulator, as long as one stays away from the charge ordered phase, we will use it in the rest of this work to investigate the real-time dynamics of photo-doped Mott insulators and of models in the metal-insulator crossover regime. 

\begin{figure*}[ht]
\includegraphics{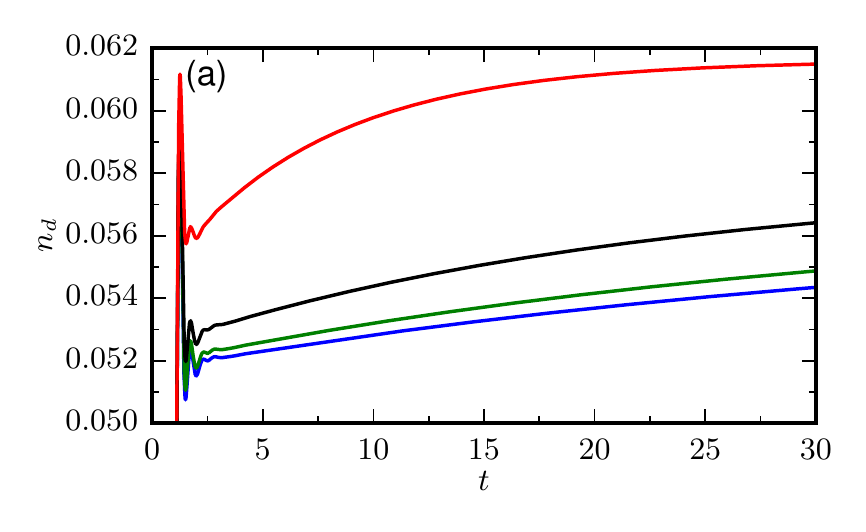}
\includegraphics[width=0.48\linewidth]{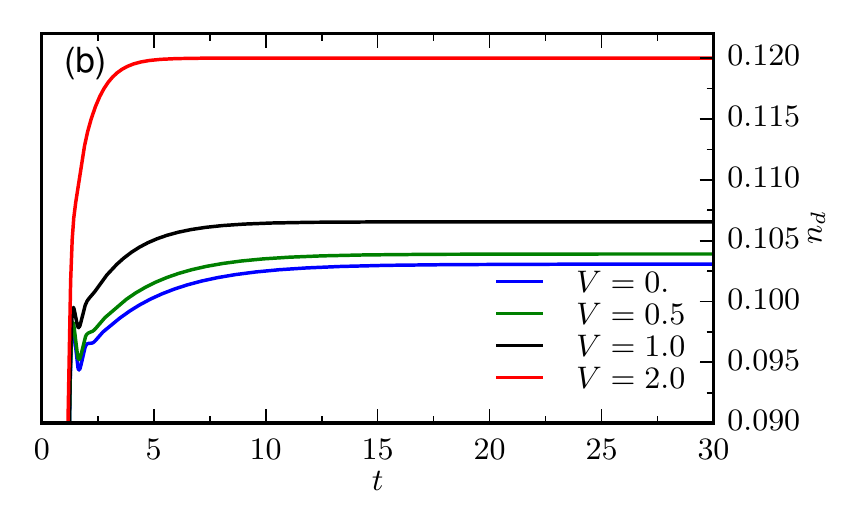}
\caption{(a),(b) Relaxation dynamics of the double occupancy $n_d$ for different values of $V$ after a pulse with frequency $\omega=10$ and pulse amplitude $E_0=5$. The on-site interaction is $U=10$ (a) and $U=7$ (b).}
\label{Fig:t_occupancy}
\end{figure*}

\section{Results}

We use the nonequilibrium EDMFT scheme to simulate the time evolution of a $U$-$V$ Hubbard model in which an electric field pulse produces a nonthermal occupation of doublons and holes. The photo-doping induced changes and the relaxation and eventual thermalization are illustrated by measuring the fermionic and bosonic spectral functions, as well as the double occupancy. To incorporate the electric field into the model, we use a gauge with pure vector potential ($\mathbf{A}$), so that the electric field is given by $\mathbf{E}=-\partial_t \mathbf{A}$.
 The vector potential enters Eq.~(\ref{Eq.:Hubbard-UV}) via the Peierls substitution, i.e. the hopping integrals $s_{ij}(t)$ acquire a time-dependent phase factor, or, equivalently, the band energies $\epsilon_\mathbf{k}$ are shifted as $\epsilon_{\mathbf{k}-\mathbf{A}(t)}$.\cite{peierls1933,davies1988} Specifically, we use a pump pulse of the form $E(t)=E_0 \sin(\omega (t-t_0)) \exp(-4.6(t-t_0)^2/t_0^2) $ with frequency $\omega,$ amplitude $E_0,$ and a Gaussian envelope. 
The width of the pulse, $t_0=2\pi/\omega$, is chosen such that the envelope accommodates a single cycle. Unless otherwise stated the pulse frequency is $\omega=8.0.$ The calculations are done for a 2D square lattice with bandwidth $W=8 |s|$ and the field is pointing along the lattice diagonal. We use $|s|$ as the unit of energy, and $\hbar/|s|$ as the unit of time.

\subsection{NCA phase diagram}
Before proceeding to the nonequilibrium dynamics we will consider the equilibrium properties in two different phases, namely the paramagnetic Mott insulating and metallic phase. In equilibrium the Mott insulator is characterized by a well defined gap in the equilibrium spectral function $A(\omega)=-\frac{1}{\pi}\text{Im}[G^{R}(\omega)]$, while the metallic phase shows a coherent quasi-particle peak, see Fig.~\ref{Fig:phase}(a). Within the NCA approximation we are limited to rather high temperatures, $T\gtrsim 1/20$. Although this temperature lies above the end-point of the Mott transition line, one nevertheless still observes a relatively sharp crossover between the metallic and insulating regimes. 
We determined the boundary between the metallic and insulating regimes by the (dis)appearance of the  quasi-particle peak in the equilibrium spectral function. This crossover line is  plotted in Fig.~\ref{Fig:phase}(b).

In order to understand the effect of the nonlocal interaction $V$ we show the spectral functions for $U=12$, deep in the Mott insulator, in Fig.~\ref{Fig:phase}(c) and for a system close to the MIT at $U=7$ in Fig.~\ref{Fig:phase}(d). In the Mott insulator, increasing the nearest neighbor coupling $V$ causes 
a slight shift of the Hubbard band toward lower frequencies and an enhanced weight of the high energy plasmon satellite, see inset of Fig.~\ref{Fig:phase}(c). Note that the high-energy peak is already present in the $V=0$ case (Hubbard model), where it is a consequence of higher order processes 
entering the strong-coupling diagrams via the hybridization function.\cite{ruhl2011}  
This can be easily  understood in the case of the Bethe lattice, where the lattice self-consistency condition simplifies to $\Delta(t,t')=|s(0)|^2 G_\text{imp}(t,t')$. The hybridization function is within NCA given by a bubble of pseudo-particle Green's functions,\cite{eckstein2010} which also leads to excitations of $2U$ from the lower Hubbard band, i.e. to a sideband at $\omega \approx -U/2+2U = 3U/2.$  The enhancement of this peak with increasing nearest neighbor interaction $V$ is a consequence of plasmonic excitations, with characteristic energy $U$, from the upper Hubbard band. These excitations contribute spectral weight at the same energy $\omega\approx U/2+U=3U/2.$ Close to the MIT the increased screening due to the inter-site interaction leads to a crossover to a coherent metallic state as a function of $V$ (cf. data for $V=2$ and $U=7$ in Fig.~\ref{Fig:phase}(d)), and a significant increase of spectral weight at high frequencies.\cite{ayral2013,huang2014,casula2012}

\subsection{Enhanced doublon relaxation due to dynamical screening}
In the following we focus on the nonequilibrium dynamics deep in the Mott insulating regime and close to the crossover line. In Fig.~\ref{Fig:t_occupancy} we plot the time evolution of the double occupancy after the pump. The amplitude of the pump is chosen such that a relatively high density of charge excitations is created in the system. (For $U=10,V=2$ the change in the double occupancy is $\Delta n_d \approx 0.04$. All the results shown in Fig.~\ref{Fig:t_occupancy} are computed for a fixed amplitude $E_0$, and thus correspond to similar values of $\Delta n_d $, because the double occupancy only weakly depends on the nearest neighbor interaction $V$.)

\begin{table}
\caption{Values of the reduced effective interaction $\mathcal{U}(\omega=0)$ measured at $t=20$ after a pulse of frequency $\omega=10$ and amplitude $E_0=5$. 
}

\begin{tabularx}{0.4\textwidth}{ XXX }
 \hhline{===}
  $V$ & $U=10$ & $U=7$ \\
  \hline
  0.5  & 9.97  & 6.92  \\

  1.0  & 9.86   & 6.67  \\

  2.0  & 9.29   & 5.47  \\
  
\end{tabularx}
\label{Table:ReducedU}
\end{table}

\begin{figure*}[ht]
\includegraphics{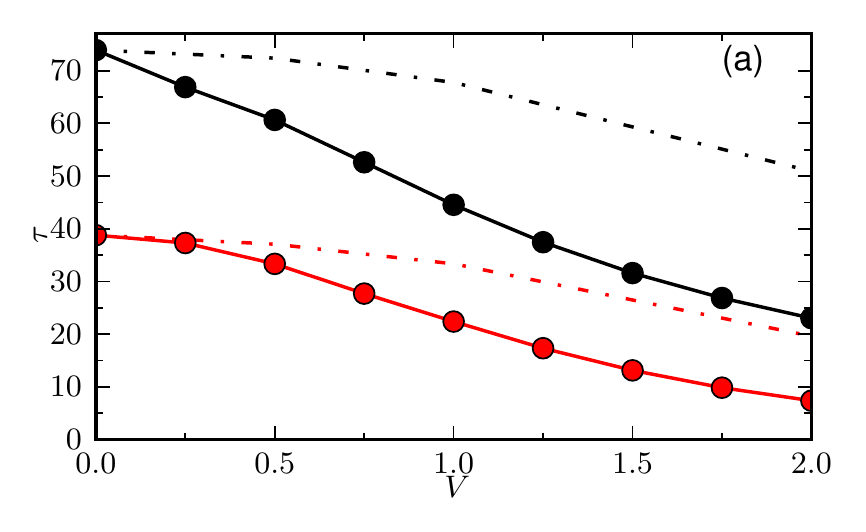}
\includegraphics{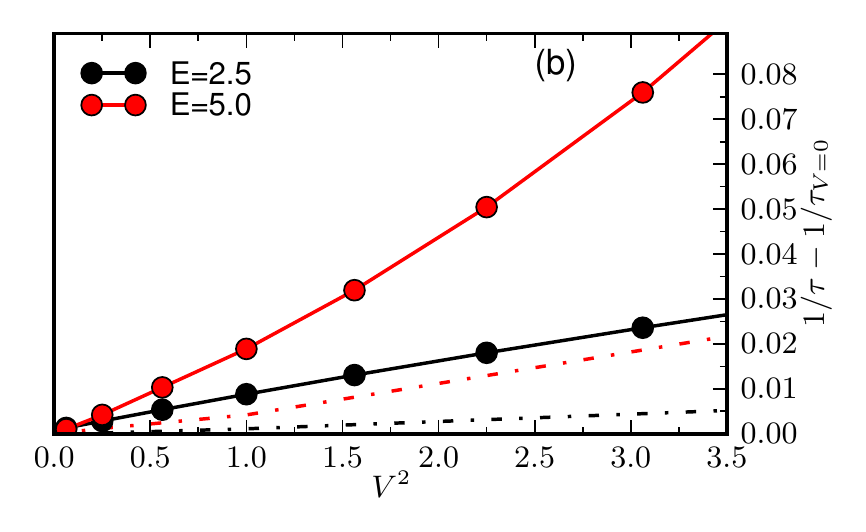}\\
\includegraphics{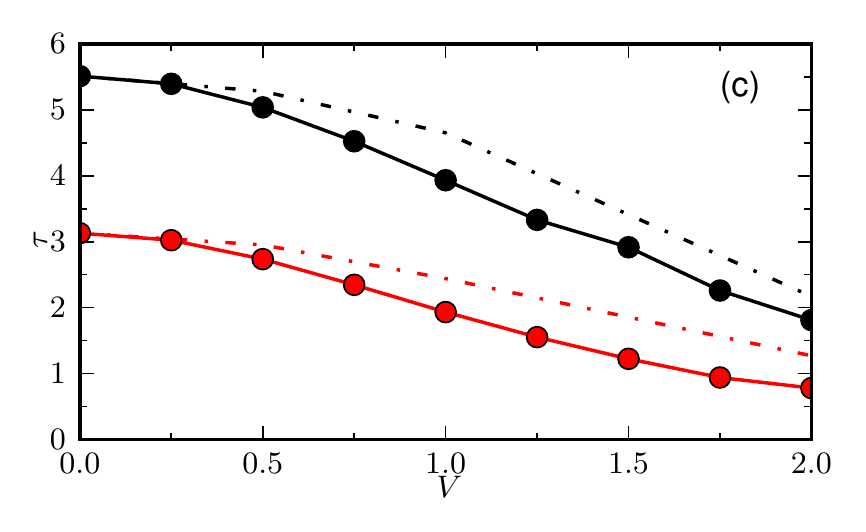}
\includegraphics{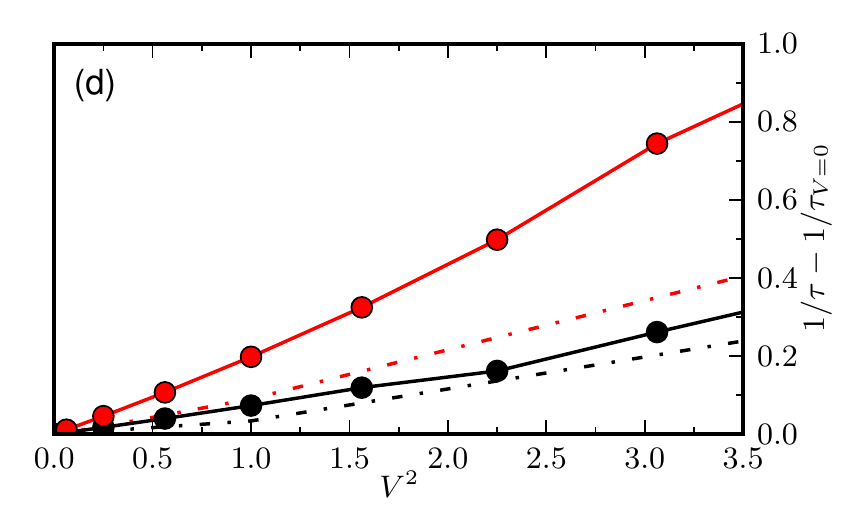}\\
\caption{
Relaxation time of the double occupancy as a function of nearest neighbour coupling $V$ for  $U=10$ (a), $U=7$ (c), pulse frequency $\omega=10$ and pulse amplitudes $E_0=2.5$ and $5.$ The dashed lines represent the relaxation time in the Hubbard model with instantaneous interaction $U=\mathcal{U}(\omega=0)$ corresponding to the static interaction in a $U$-$V$ Hubbard model with a given $V$ (see discussion in the main text). (c,d) Scaling of the difference between the inverse relaxation times at $V=0$ and $V>0$ with $V^2$ for the same parameters as in panels (a,c).}
\label{Fig:relax_time}
\end{figure*}

After a transient dynamics the double occupancy follows an exponential relaxation. In the Hubbard model, the relaxation time increases exponentially with increasing $U$ if the gap size is large. \cite{eckstein2013,sensarma2010} In small gap insulators, impact ionization processes can lead to a rapid carrier multiplication at short times, followed by a slower exponential relaxation.\cite{werner2014} 
As can be seen from the equilibrium spectra plotted in Fig.~\ref{Fig:phase}(a), even $U=10$ lies within this small gap regime,  so that we will focus on the long-time relaxation. 
The relaxation dynamics for different values of the nearest neighbour interaction $V$ are plotted in panels (a) and (b) for $U=7$ and $U=10$, respectively. We fit the relaxation by an exponential function $n_d(t)=n_d(t=\infty)+A\exp(-t/\tau)$ in the range $5<t<30$ to extract the relaxation times.
While the relaxation curves look qualitatively similar, the relaxation times, plotted in Fig.~\ref{Fig:relax_time}(a) and (c), are strongly reduced for larger $V$. 

\begin{figure*}[ht]
\includegraphics{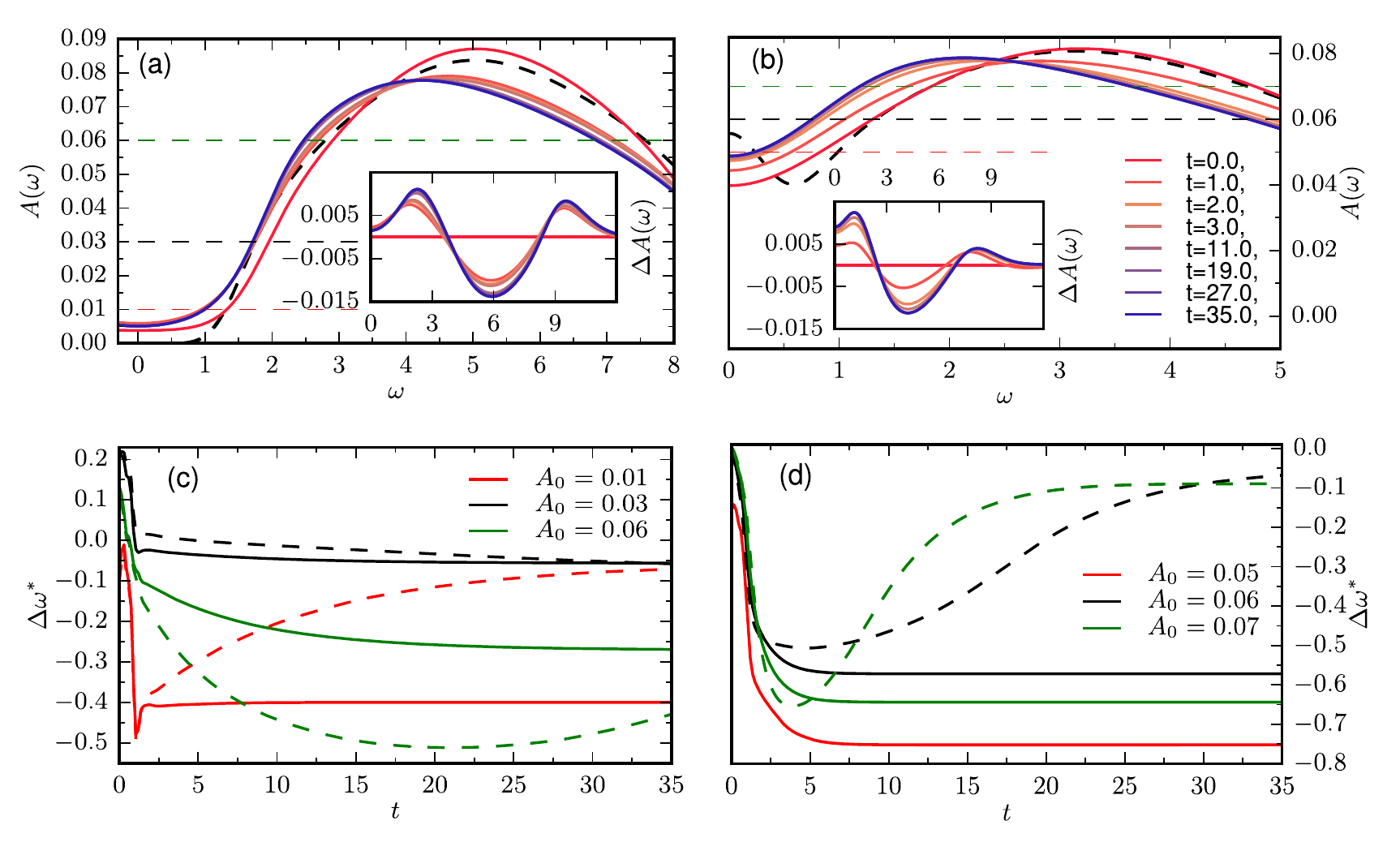}
\caption{(a,b) Partial Fourier transform of the spectral function $A(\omega,t)$ with insets showing the difference from the initial value $A(\omega,t)-A(\omega,t=0)$ for $U=10,V=2,E_0=5.0$ (panel (a)) and $U=7,V=2,E_0=3.5$ (panel (b)). The dashed black lines represent the initial equilibrium spectral function. (c,d) Change  in the gap size measured by the frequency $\omega^*$ at which the spectral function is equal to some fixed value, $A(\omega^*)=A_0$, for $U=10$ (c) and $U=7$ (d). The dashed lines represent the $\omega^*$ for the same parameters but with coupling to an external bath with $\lambda=1,$ see also the discussion in the main text. For $U=7$, there is no intersection for $A_0=0.05$ in the case of coupling to external bath. The corresponding cuts through the spectral function are shown by the dashed lines in panels (a,b).
}
\label{Fig:spectral}
\end{figure*}

Several effects could potentially explain this observation. On the one hand, 
the dynamical screening reduces the effective instantaneous interaction $\mathcal{U}(\omega=0)$ (see Sec.~\ref{Sec.:eff_interac} for a detailed analysis), so the dynamics may be more properly described by a Hubbard model with an effectively reduced repulsion $U=\mathcal{U}(\omega=0),$ which in turn leads to a faster relaxation. In order to test this scenario we measured the screened effective interaction $\mathcal{U}(\omega=0)$ after the photo-doping pulse for different values of the nearest neighbor interaction $V$, see Table~\ref{Table:ReducedU}, and then recalculated the dynamics in the Hubbard model ($V=0$) using these reduced interaction parameters. The corresponding relaxation times are shown by dashed lines in Fig.~\ref{Fig:relax_time}(a) and (c).
At $U=7$, in the metal-insulator crossover regime, the reduction in the effective on-site interaction (and hence gap-size) contributes substantially to the faster relaxation times, but in the Mott insulator case ($U=10$), this effect cannot explain the observed large changes.

The second scenario is that the relaxation rate is enhanced due to the absorbtion of bosonic collective excitations (plasmons), which opens an additional relaxation channel. The coupling to the bosonic excitations can be understood from the form of the electron-boson action in Eq.~(\ref{Eq.:Action_EDMFTwBoson}), which is equivalent to the action of a Anderson-Holstein model with a coupling to a continuum of free bosons.\cite{ayral2013} Previous studies of the relaxation dynamics in the Hubbard-Holstein model\cite{werner2013, werner2015} showed an enhanced relaxation of the charge excitations due to the coupling with phonons. In order to check the latter idea we can use the empirical Matthiessen's rule $1/\tau=1/\tau_{V=0}+1/\tau_b$ to separate the inverse relaxation time into an electronic and bosonic contribution.  Based on this formula we extract the inverse relaxation time due to the scattering with the bosonic bath $1/\tau_b$ from the relaxation times obtained by fitting. The result is presented in Fig.~\ref{Fig:relax_time}(b,d).

We note that the extracted relaxation times $1/\tau_b$ 
are proportional to $V^2$ for small $V$. One can understand this scaling from the effective Hamiltonian representation of the impurity model, which corresponds to an Anderson impurity model coupled to the bath of bosonic modes.\cite{ayral2013,hafferman2014} The relaxation time due to the coupling to the bosonic bath can be approximated by the Fermi golden rule:
\bsplit{
  \frac{1}{\tau_b} &\approx \sum_k \lambda_k^2 A(\omega-\omega_k)\approx A(\omega_F) \sum_k \lambda_k^2 =\\
   &- \frac{A(\omega_F)}{\pi} \int d\omega \text{Im}[\mathcal{D}(\omega)]
   \propto A(\omega_F) V^2,
}
where $\lambda_k,\omega_k$ are the coupling constants and phonon frequencies for the $k$-th bosonic mode and $A(\omega)$ is the electronic spectral function. The main simplification used was to approximate the spectral function at the final energy $A(\omega_F=\omega-\omega_k)$ as a constant. The numerical investigation in Ref.~\onlinecite{ayral2013} showed that for weak nearest neighbour interaction $V$ the integral over the bosonic Weiss field scales as $\int d\omega \text{Im}[\mathcal{D}(\omega)]\propto V^2.$ Hence, our phenomenological analysis shows that the enhancement  of the relaxation due to the nonlocal interaction $V$ and the corresponding $V^2$ scaling is consistent with a relaxation aided by the coupling to a bath of bosonic degrees of freedom (plasmons).\cite{murakami2015,golez2012,werner2013,dorfner2015,sayyad2015}

\subsection{Spectral properties}\label{Sec.:spectral}

Further insight into the relaxation dynamics is obtained from the evolution of the partially Fourier transformed spectral function 
$A(t,\omega)=-\frac{1}{\pi}\text{Im}\int_{t}^{t+t_\text{max}} dt' e^{-\I \omega (t'-t) }G^{R}(t',t),$ 
where we use $t_\text{max}=10.$ First we study the effect of the screening in the Mott insulating phase ($U=10$), where the equilibrium spectral function consists of the upper and lower Hubbard band separated by a well defined gap (Fig.~\ref{Fig:spectral}a). The pulse leads to an increase in the number of charge carriers (doublons and holes), which in turn results in stronger screening. The spectral weight at $\omega=0$ increases slightly 
due to the effect of screening, heating and doping. 
The enhanced screening effect manifests itself on the timescale of $1/\text{bandwidth}$ (note the nonuniform time mesh in Fig.~\ref{Fig:spectral}). 
At longer times the screening is further increased due to the redestribution of the charge carriers within the upper Hubbard band and on even longer time scales due to 
doublon production associated with thermalization. 

In order to compare these results to the evolution of a system close to the metal-insulator crossover ($U=7$), we adjust the pulse strength $E_0$ such that the same amount of photo-doped charge carriers $\delta n_d \approx 0.04$ is present after the excitation. The result is shown in Fig.~\ref{Fig:spectral}(b). While the quasi-particle peak quickly disappears there is again a shift and broadening of the Hubbard bands, associated with an increase in spectral weight in the pseudo-gap region. The insets in Fig.~\ref{Fig:spectral}(a),(b) show the evolution of the difference of the time-dependent spectral function $A(\omega,t)-A(\omega,t=0)$ for $\omega>0$. The spectral weight at the lower and upper edge of the Hubbard band is increased and as a consequence the gap is reduced and partially filled in. Since the system is almost thermalized for $t>10$ these changes can be attributed to the heating of the system, which we have confirmed by comparison to the equilibrium spectral functions at elevated temperatures.

In particular, the increase of weight at low $\omega$ is related to the partial filling-in of the gap, while the increase at large $\omega$ is related to the appearance of side-bands in the photo-doped or thermally excited system (See Fig.~\ref{Fig:phase}(c,d)). 

In order to analyze the doping and screening induced changes in the gap size we plot the evolution of the frequency $\omega^*$, where the spectral function takes some fixed value: $A(\omega^*)=A_0$, see Fig.~\ref{Fig:spectral}(c,d). The corresponding cuts  are shown as dashed lines in Fig.~\ref{Fig:spectral}(a,b). At the shortest times the charge carriers are inserted near the middle of the upper Hubbard band, which results in a distortion of the spectral function (the dashed black line shows the initial equilibrium result). 
While the pulse is on ($t\lesssim 3$) there are field induced effects, which lead to a nontrivial dynamics of the spectral function. 
The doping, heating and screening induced changes result in an asymmetric reshaping of the Hubbard bands (see insets)  and the modifications in the gap size are quite different from what one would obtain from a rigid shift of the Hubbard bands of the undoped system. 
In order to investigate the reduction of the gap quantitatively we analyze the difference of the frequency $\omega^{*}$ from the equilibrium value, $\Delta \omega^{*}(t)=\omega^{*}(t)-\omega^{*}_{eq}.$ The rapid reduction of $\Delta \omega^{*}$ just after the excitation is related to the fast reduction of the effective $U$ due to screening, see Sec.~\ref{Sec.:eff_interac}. The longer time dynamics depends on the gap size: for $U=10$ the gap keeps decreasing even on the longest accessible times scales, while for $U=7$ the gap size rapidly stabilizes. This difference can be explained by the longer thermalization time in the $U=10$ case.

 The dashed lines in Fig.~\ref{Fig:spectral}(c,d) show the 
 analogous results for a model with local coupling to a heat bath; 
 for details and further analysis see Sec.\ref{Sec.:Bath}. Due to the cooling by the heat bath, 
 the gap recovers its original value, as will be 
further discussed in Sec. \ref{Sec.:gap_reduction}.


\begin{figure*}[ht]
\includegraphics[scale=0.95]{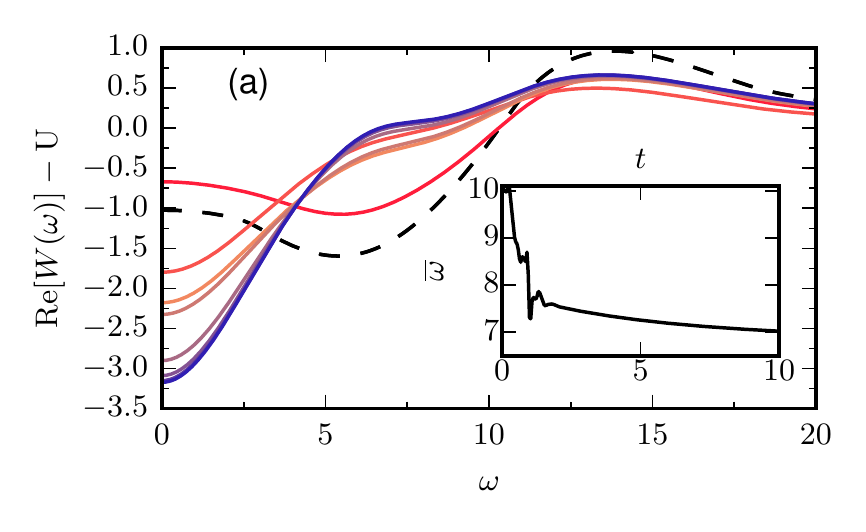}
\includegraphics[scale=0.95]{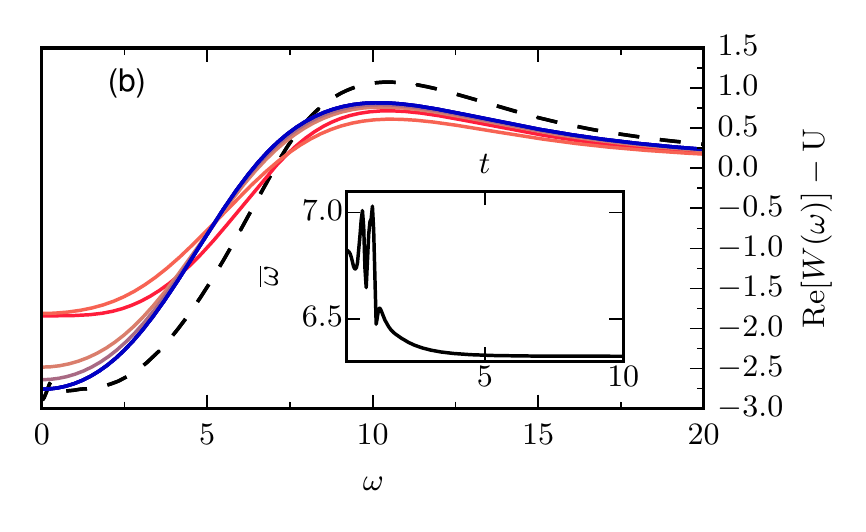} \\
\includegraphics[scale=0.95]{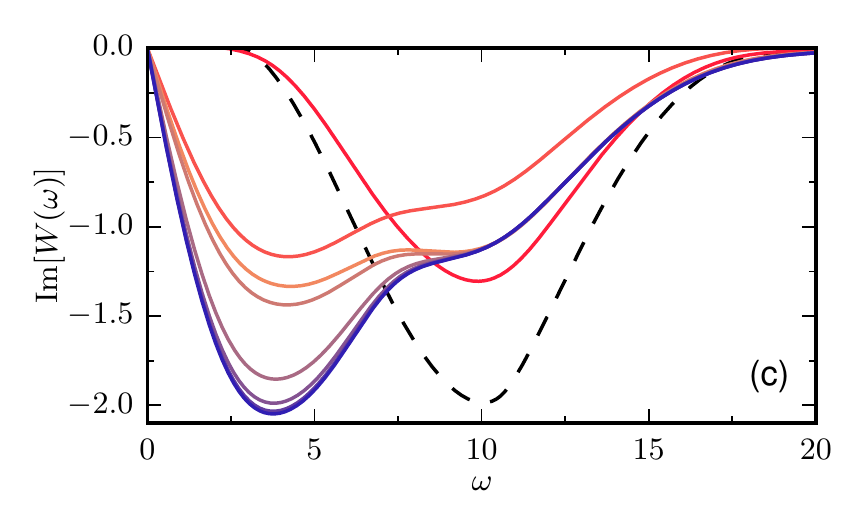}
\includegraphics[scale=0.95]{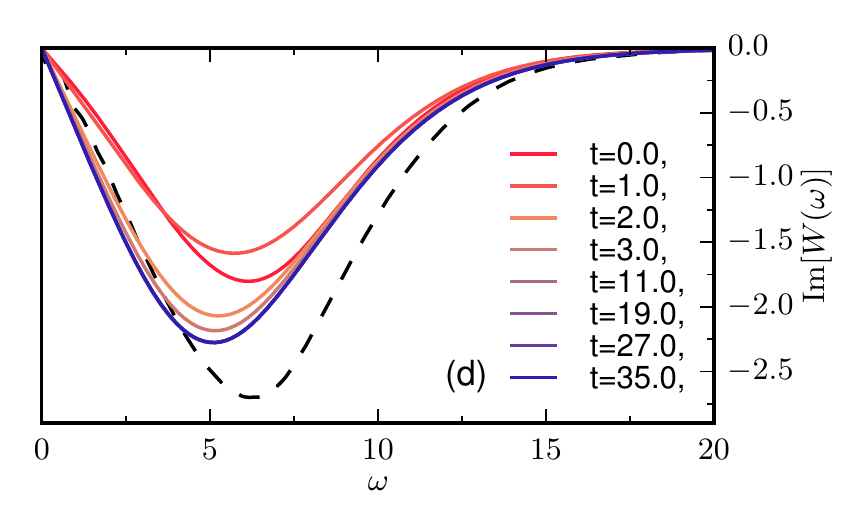} \\
\includegraphics[scale=0.95]{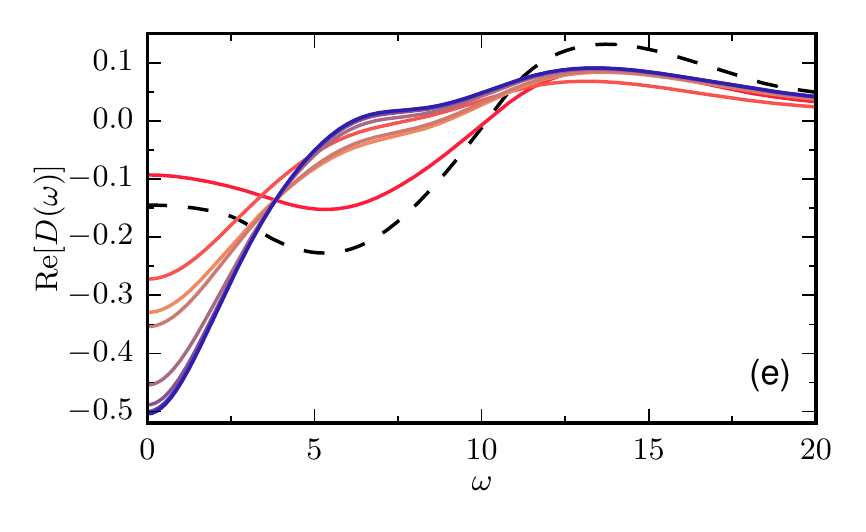}
\includegraphics[scale=0.95]{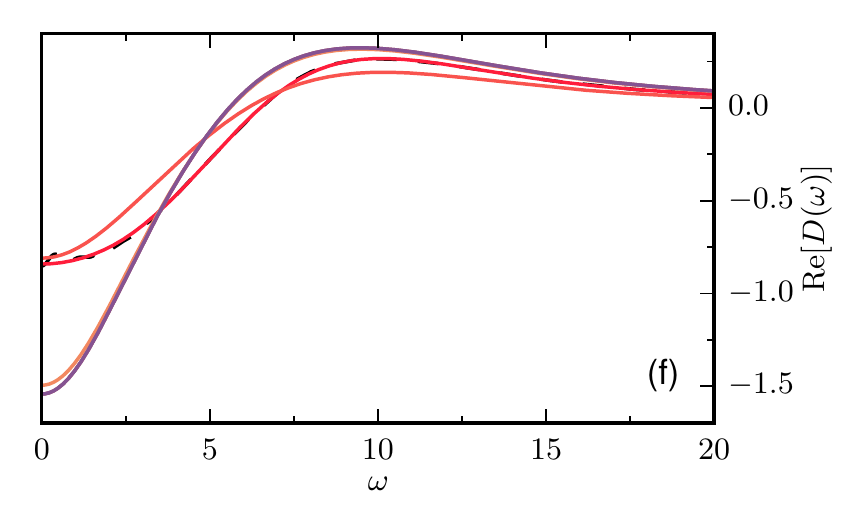} \\
\caption{Real (a,b) and imaginary (c,d) part of the partial Fourier transform of the screened interaction $\text{Re}[W(t,\omega)]-U$ for $U=10,E_0=5$ (left column) and $U=7,E_0=3.0$ (right column) at fixed $V=2,\omega=8.$ The insets in (a,b) show the evolution of the average boson frequency $\overline \omega(t)$. The black dashed lines represent the screened interaction in the initial equilibrium state. (e,f) Real part of the partial Fourier transform of the regular part of the partially screened on-site interaction $\text{Re}[\mathcal{D}^\text{reg}(t,\omega)]=\text{Re}[\mathcal{U}(t,\omega)]-U$  for $U=10,E_0=5$ (e) and $U=7,E_0=3.5$ (f). }
\label{Fig:W_t}
\end{figure*}

\subsection{Effective interaction}\label{Sec.:eff_interac}
Within EDMFT, the inter-site interaction translates into a retardation of the effective on-site interaction $\mathcal{U}(t,t'),$ while the fully screened interaction $W(t,t')$ also includes the local screening effects. In equilibrium and in the Mott insulating phase the imaginary part of the fully screened interaction $\text{Im}[W(\omega)]$ consist of one broad peak around $\omega \approx U,$ which is weakly shifted to lower frequencies upon increasing $V$.\cite{ayral2013,huang2014} In the chemically doped Mott insulator, or in the strongly correlated metal phase, $\text{Im}[W(\omega)]$ exhibits a two-peak structure. The low energy peak is associated with transitions between quasi-particle peak and Hubbard bands, while the higher energy peak at $\omega \approx U$ is associated with transitions between the Hubbard bands. 
\begin{figure*}[ht]
\includegraphics{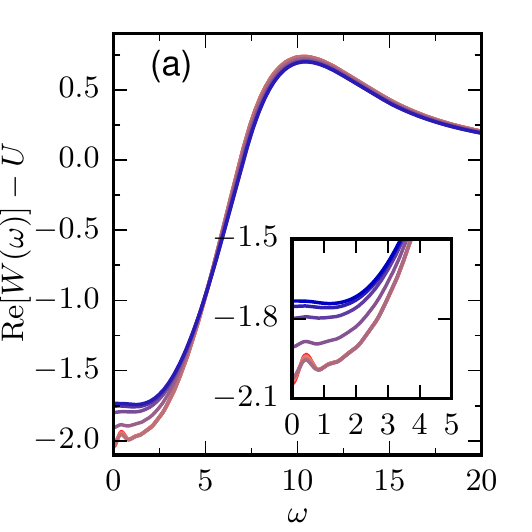}
\includegraphics{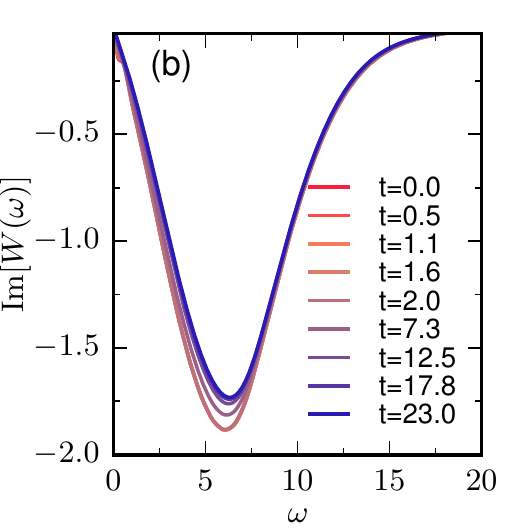}
\includegraphics{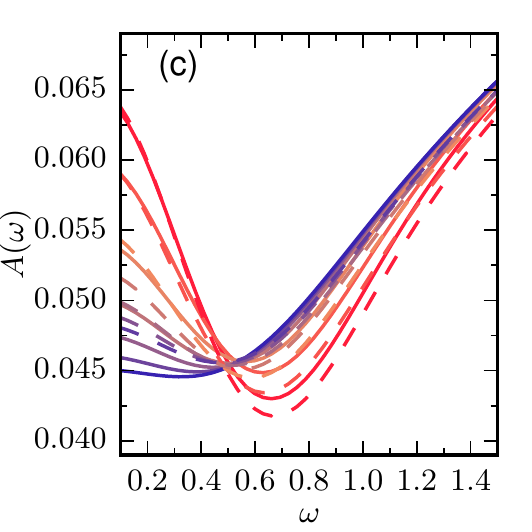}
\caption{Real (a) and the imaginary (b) part of the partial Fourier transform of the screened interaction $W(t,\omega)$ for $U=6.9$, $V=2$, $E=0.25$, and $\omega=2.$ (c) Comparison of the dynamics of the partial Fourier transform of the spectral function $A(\omega,t)$ for the $U$-$V$ (full lines) model and the Hubbard model (dashed lines) with a correspondingly reduced on-site interaction $U=6.425.$}
\label{Fig:W_increase}
\end{figure*}
By increasing the doping, the weight of the low energy peaks is strongly enhanced. 
The real part of the fully screened interaction $\text{Re}[W(\omega)]$ is reduced for low frequencies as we increase the nearest neighbour interaction $V$ or the doping, reflecting the stronger screening effect. 

An interesting question is how these spectral functions change after a photo-doping excitation in a Mott insulator. In order to get insights into the nonequilibrium dynamics of screening we perform the partial Fourier transform of the fully screened interaction to obtain 
$W^{R}(t,\omega)=\int_{t}^{t+t_\text{max}}dt' e^{-\I \omega (t'-t) }W^{R}(t',t).$
The results for $U=10$, $V=2.0$ and a photo-doping concentration of $\Delta n_d=0.04$ are shown in Fig.~\ref{Fig:W_t} (a),(c),(e). In equilibrium the imaginary part of $W$ shows a broad peak at $\omega_p \approx U,$ while the real part is slightly reduced for $\omega<\omega_p$ as a result of local screening. During and shortly after the pulse ($t\lesssim 3$) there is a strong 
modification in the distribution and coupling strength of the screening modes. 
Due to the photo-doping and the screening from low-energy excitations within the Hubbard bands, a broad low-energy peak appears in the imaginary part of $W^R(t,\omega)$ and the instantaneous effective interaction $\text{Re}[W^R(t,\omega=0)]$ is strongly reduced.

Even though the screened interaction $W(\omega)$ of the photo-doped Mott insulator looks qualitatively similar to that of a chemically doped Mott insulator, the origin of the low-energy peak in $\text{Im}[W^R(\omega)]$  is different: The single-particle spectral function of photo-doped Mott insulators does not feature a quasi-particle peak at $\omega=0$, nor sharply defined quasi-particle features near the gap edges.\cite{eckstein2013} The low-energy peak in $\text{Im}[W^R(\omega)]$ is therefore not associated with transitions between a well-defined quasi-particle band and the Hubbard bands, but rather with excitations within the Hubbard bands. As a result, the feature is broader in the photo-doped system than in a chemically doped one. 

A useful quantity to characterize the screening is the average boson frequency 
$$\overline \omega(t)=\frac{\int_0^{\infty} d\omega ~\omega ~ \text{Im}[W(t,\omega)]}{ \int_0^{\infty} d\omega ~ \text{Im}[W(t,\omega)] }.$$  
It strongly decreases during the pulse ($t \lesssim 2.5$), as shown in the inset of Fig.~\ref{Fig:W_t}(a,b). In the case of $U=10$ the initial fast drop of $\overline \omega(t)$, which is a consequence of the doping-induced appearance of the low energy mode, is followed by a slower long time relaxation associated with changes in the energy distribution of the photo-carriers. For $U=7$, the system is essentially thermalized after $t\approx 7$ and no further changes in the bosonic spectral function or average screening frequency occur. 

We note that the changes in the screened interaction are less dramatic for $U=7$ than for $U=10$, see Fig.~\ref{Fig:W_t}(a,b). This can be understood by the fact that the initial state is metallic, and already has low-energy screening modes. The heating destroys the quasi-particle peak (i.e the system moves across the metal-insulator line in temperature), so that the final state is a thermally excited insulator for which the screening is not much larger than in the metallic initial state. In contrast, for $U=10$ the initial and final states correspond to a cold and hot Mott insulator with a very different number of thermally excited carriers. 
The partial Fourier transform of the regular part of the effective on-site interaction $D(t,t')=\mathcal{U}(t,t')-\delta_c(t,t')U $ shows a behavior which is qualitatively similar to that of the screened interaction $W(t,t').$ The main difference is that the reduction is smaller (see Fig.~\ref{Fig:W_t}(e,f)), since the local screening effects are absent. The time-dependent  changes in the effective interactions are consistent with the observed changes in the spectral function, where a strong reduction of the gap size is observed on the time scale $1/\text{bandwidth}$.

Photo-doping a metallic state destroys the quasi-particle peak,\cite{eckstein2013} which leads to a loss of low-energy excitations and hence low-energy screening. This effect competes with the increase of the screening due to the photo-doping. In a strongly excited system, the two effects cannot be easily disentangled. To single out the effect of the destruction of the quasi-particle peak, we choose a low-frequency and low amplitude pulse, namely $\omega_0=2.0$ and $E_0=0.25$, which reduces the photo-doping effect. With this pulse we indeed observe a decrease of the screening in the low energy regime $\omega\leq2$ of the real part of the partial Fourier transform of the fully screened interaction $W(t,\omega)$, see Fig.~\ref{Fig:W_increase}. In addition to this increase of the real part of $W$, the low-energy feature associated with excitations within the quasi-particle band disappears. The 
latter excitations are 
responsible for the pole-like structure near $\omega=0.3$, which disappears rapidly in agreement with the dynamics of the spectral function, see Fig.~\ref{Fig:W_increase}(c) and inset of (a).

The reduced screening effect feeds back onto the spectral function, and leads to a further decrease of spectral weight in the low-energy region in comparison with the dynamics of the Hubbard model with appropriately reduced $U$. For the comparison in panel (c), we have chosen $U$ such that the equilibrium spectral function reproduces the result of the $U$-$V$ Hubbard model at low-energies ($\omega \leq 2$) in the initial state. As can be seen,  the reduction of the spectral weight is more pronounced in the case of the $U$-$V$ model due to the change in the  screening environment. 

\section{Coupling to a thermal bath}
\label{Sec.:Bath}
To describe the dissipation of energy to external degrees of freedom we need to couple the system to some environment.\cite{aoki2014_rev} On short times, the dissipative environment will lead to a redistribution of the energy of the photo-doped carriers (intra-band relaxation), while on longer timescales the bath enhances recombination processes, and enables the system to relax back towards the initial equilibrium state. Here we will in particular study how the  screening is influenced by the transient modification of the energy distribution of the photo-doped carriers.

Technically, one can integrate out the environment to obtain an effective description of the system by adding a corresponding contribution to the electronic self-energy $\Sigma$. We use the lowest order diagram for a Holstein like electron-phonon coupling,\cite{eckstein2013} namely $\Sigma_\text{diss}(\omega)=\lambda^2 G(t,t') B_0(t,t'),$ where $\lambda$ is the coupling strength and $B_0(t,t')$ is the free bosonic propagator with a linear spectral density between a low-energy and high-energy cutoff, namely $-\frac{1}{\pi} \text{Im}[B_0(\omega)]=\omega $, if  $ \omega_l<\omega< \omega_h$ with $\omega_l=0.2$ and $\omega_h=1.0$, and zero elsewhere. The bath temperature is set to $T=1/20.$ Since this treatment is only suitable in the weak coupling regime we will restrict ourselves to values of the coupling ($\lambda < 1.0$), for which the bath only results in a slight broadening of the spectral function in equilibrium. 
The spectral distribution of the bath has a substantial effect on the relaxation dynamics. To efficiently relax the states around the quasi-particle peak the system needs to be able to excite low energy bosons, while a restriction to only low energy bosons will slow down the relaxation of high-energy carriers due to the necessity for high order processes. 
For this reason, we choose a continuous spectrum which includes low as well as relatively high energy bosons. 

\subsection{Relaxation with bath}

\begin{figure*}[ht]
\includegraphics{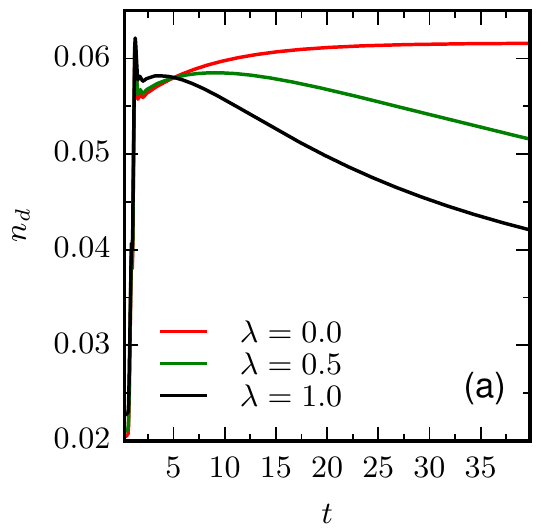}
\includegraphics{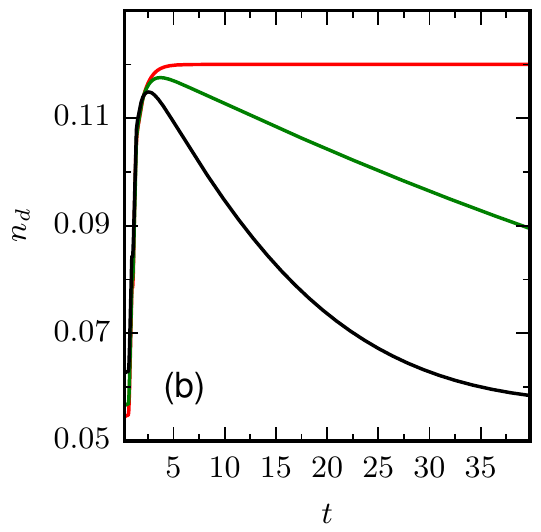}
\includegraphics{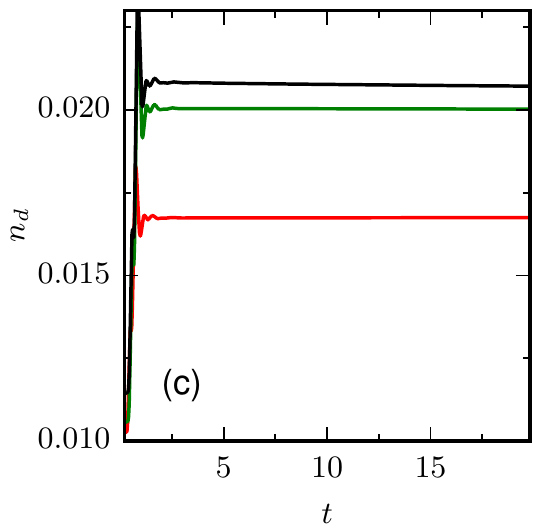} 
\\
\includegraphics{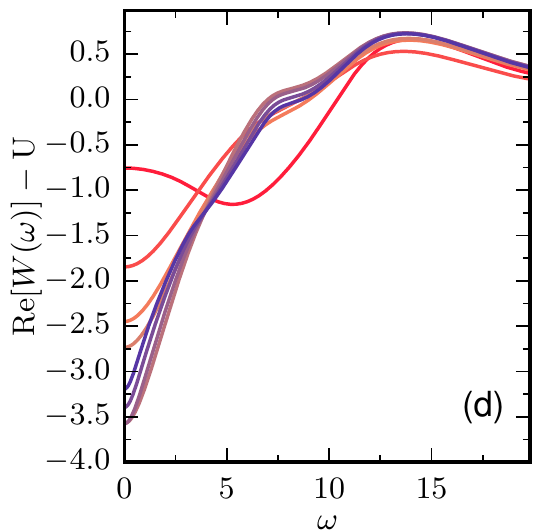}
\includegraphics{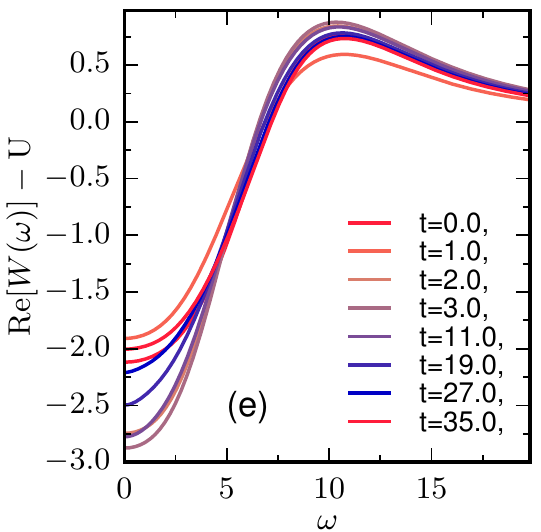}
\includegraphics{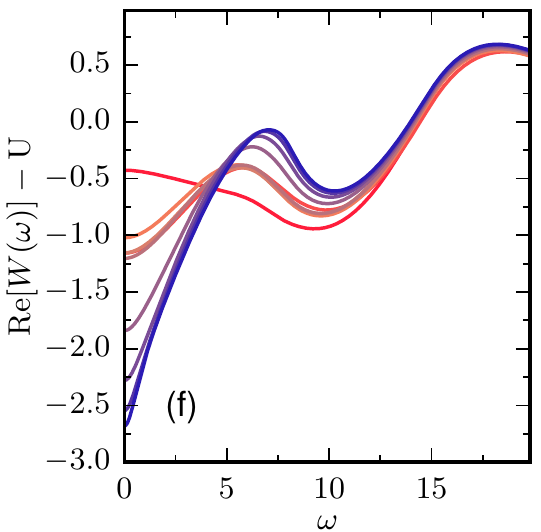}\\
\caption{ 
Relaxation dynamics of the double occupancy $n_d$ for $U=7$ (a), $U=10$ (b), and $U=14$ (c) with
different values of the coupling to the bath $\lambda$ during and after the pulse with frequency $\omega=8$ and pulse amplitude $E_0=5.$
The real part of the partial Fourier transform of the screened interaction $\text{Re}[W^\text{reg}(t,\omega)]=\text{Re}[W(t,\omega)]-U$ for  $U=7$ (d), $U=10$ (e) and $U=14$ (f) at fixed $V=2.0,\lambda=1.0,E_0=5.$}
\label{Fig:nd_bath}
\end{figure*}

In contrast to the system without bath, which is approaching a thermal equilibrium state at higher temperatures, the addition of the heat bath ensures that the system relaxes back to the initial state, as can be seen in the evolution of the double occupancy in Fig.~\ref{Fig:nd_bath}(a,b). This behavior is consistent with the previous investigation of the Hubbard model.\cite{eckstein2013} The question which we would like to address here is the effect of the bath on the screening.

For the Mott insulator case $U=10$ the bath tends to reduce the screened interaction $\text{Re}[W(\omega)]$ at low frequencies on the observed time scale. This is a consequence of the faster relaxation of the highly excited doublons to the lower edge of the Hubbard band, which enhances the screening, compare Figs.~\ref{Fig:nd_bath}(c) and \ref{Fig:W_t}(a). 
The increase in $\text{Re}W(\omega)$ at later times is the result of the bath-enhanced doublon-hole recombination.  
Deeper in the Mott phase ($U=14$), the recombination is suppressed and even with a coupling to a heat bath the double occupation remains approximately constant on the accessible time scales (not show). 
The bath nevertheless enhances the relaxation of the high energy doublons to the lower edge of the upper Hubbard band, which leads to 
monotonically increasing screening effect with time, 
see Fig.~\ref{Fig:nd_bath}(e). 
For $U=7$, the screening dynamics is non-monotonous: the initial increase of charge carriers enhances the screening, but the coupling to the bath leads to a faster recombination of the charge carriers, which eventually reduces $W(\omega,t)$ to a function close to the initial equilibrium screened interaction, see Fig.~\ref{Fig:nd_bath}(d). 

\subsection{Photo-doping in the metal-insulator crossover regime}

An appealing scenario would be the appearance of a transient metallic state (with quasi-particle peak) after the photo-doping of an insulating initial state, as a result of enhanced screening. To investigate this possibility, we have systematically studied the photo-induced dynamics for a system close to the metal-insulator crossover line, where this effect may be expected to occur.

 
We found that photo-doping an insulator in the vicinity of the metal-insulator crossover line can trigger a nontrivial evolution which reflects the effects of increased scattering and of the changes in the screening environment, but we were not able to realize a screening-induced metal state even close to the metal-insulator regime, where it is easy to realize a set-up in which $\mathcal{U}(\omega=0)$ drops below the critical static value for which one finds a metallic solution in equilibrium. A potential reason may be the detrimental effect of doublon/hole scattering on the emergence of coherent quasi-particles. This resembles the observation that the build-up of a coherent quasi-particle peak in a photo-doped Mott-insulator  takes at least longer than the quasi-particle lifetime.\cite{eckstein2013} We also used rather weak pulses, since  stronger pulses enhance the destruction of the quasi-particle peak and in addition lead to stronger heating. Another limitation is the NCA based impurity solver, which systematically underestimates the metallic nature of the solution, and which does not allow us to study the low-temperature behavior in the vicinity of the first-order metal-insulator transition.

\begin{figure*}
\includegraphics{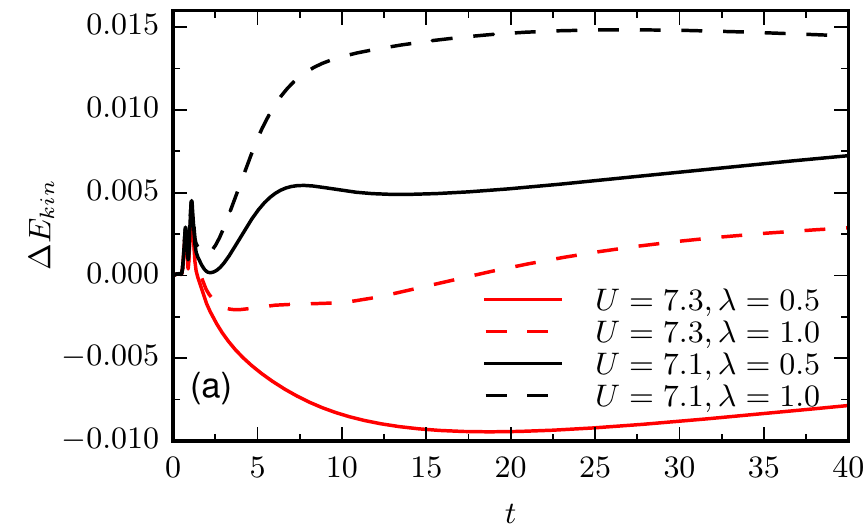}
\includegraphics{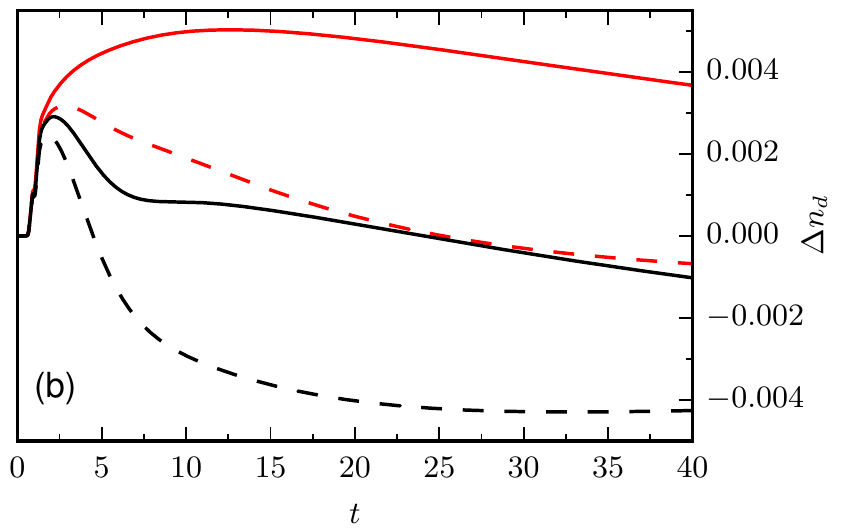}  \\
\includegraphics{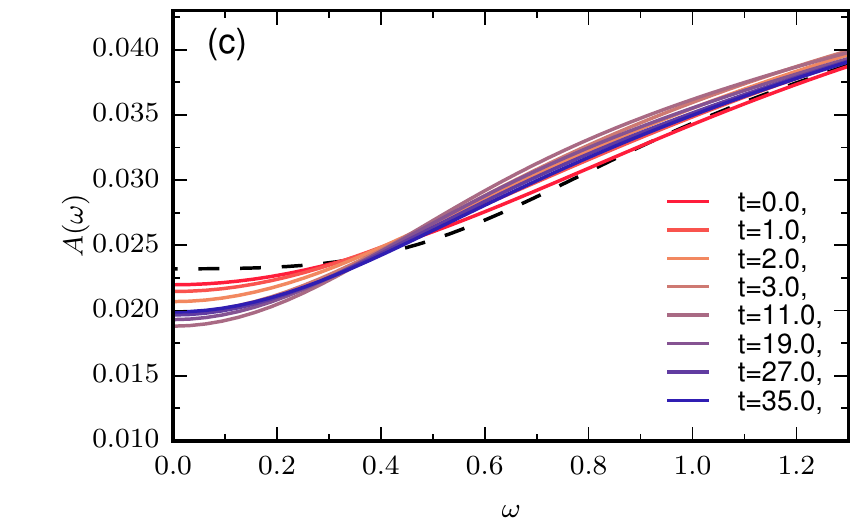}
\includegraphics{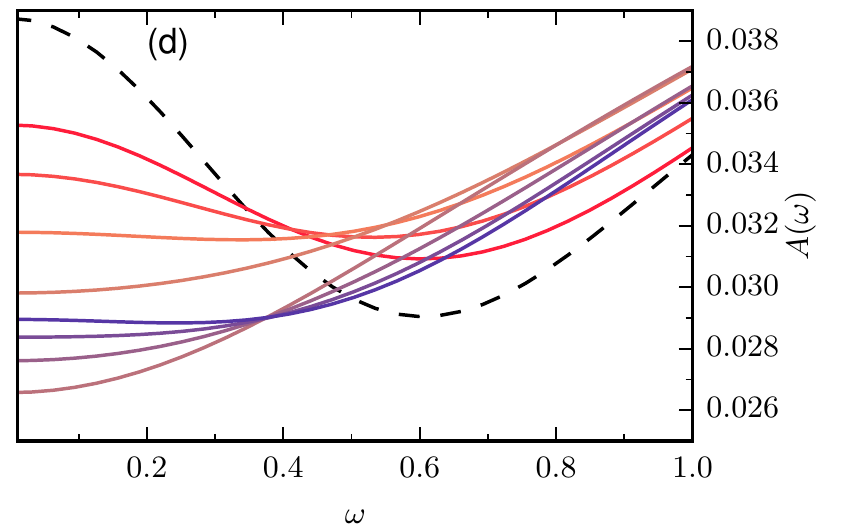} \\
\includegraphics{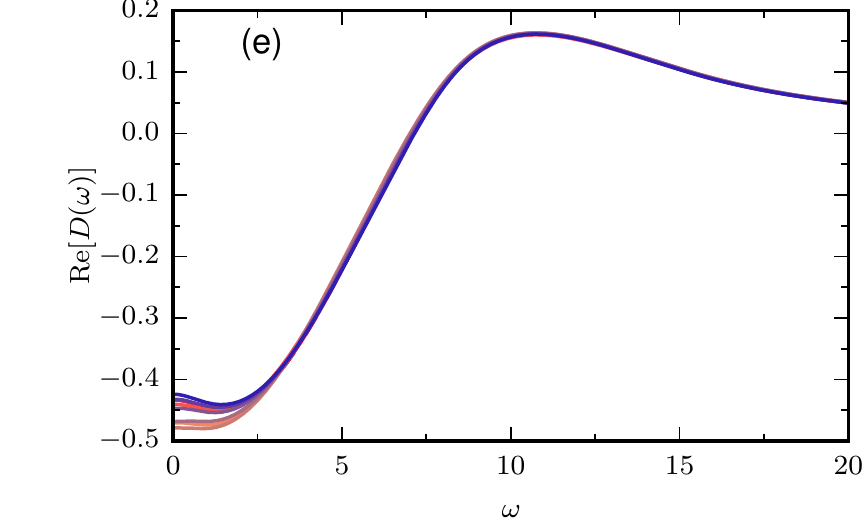} 
\includegraphics{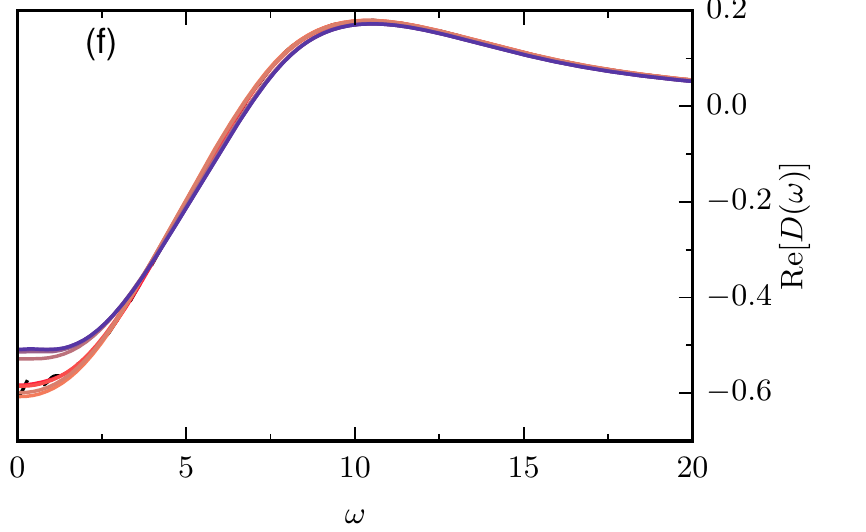} \\
\includegraphics{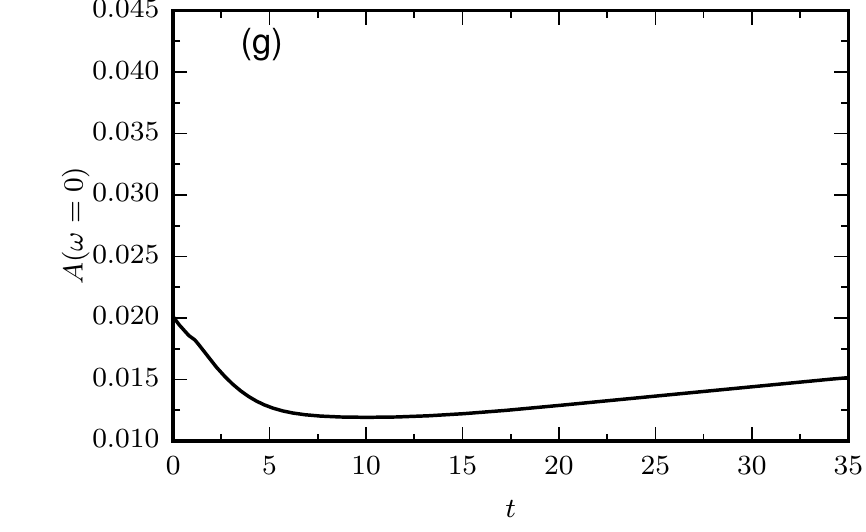}
\includegraphics{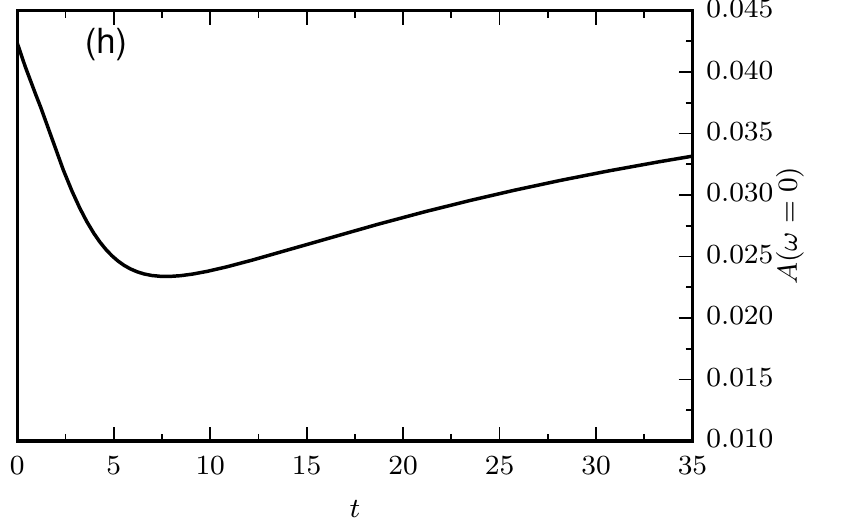}
\caption{Relaxation dynamics of the kinetic energy $E_\text{kin}$ (a) and the double occupancy $n_d$ (b) for $U=7.3$ and $U=7.1$ at $V=2.0$ for different values of the coupling with the bath $\lambda=0.5$, $1.0$ during and after a pulse with $\omega=8$ and $E_0=1.0$. The partial Fourier transform of the spectral function $A(\omega,t)$ for  $U=7.3$ and  $U=7.1$ is shown in panels (c) and (d). 
The dashed lines corresponds to the equilibrium spectral function. 
Panels (e,f) plot the real part of the partial Fourier transform of the regular part of the partially screened on-site interaction $\text{Re}[\mathcal{D}(t,\omega)]=\text{Re}[\mathcal{U}(t,\omega)]-U$ for $U=7.3$ (e) and $U=7.1$ (f). The time evolution of the spectral weight $A(\omega=0,t)$ for $U=7.3$ (g) and  $U=7.1$ (h) for $\lambda=1$.}
\label{Fig:MIT_bath}
\end{figure*}

Photo-doping in the vicinity of the metal-insulator crossover line can nevertheless trigger a nontrivial evolution which reflects the effects of increased scattering and of the changes in the screening environment. Figure~\ref{Fig:MIT_bath} shows the change in the kinetic energy and in the double occupation after a weak pulse in systems with weak and strong coupling to a heat-bath. We compare the system at $U=7.1$, where in equilibrium a weak quasi-particle peak is present and at $U=7.3$, where the system only exhibits a pseudogap. Let us first consider the insulating system with weak coupling to a heat-bath ($\lambda=0.5$). After the pulse, the kinetic energy drops, while the doublon number increases, as expected in a photo-doped insulator with dissipation. At longer times, the recombination of doublon-hole pairs leads to a slow relaxation back to the initial equilibrium state. The system with strong coupling to a heat bath ($\lambda=1$) shows a more puzzling behavior. After an initial drop, the kinetic energy starts to increase beyond the initial equilibrium value, while the double occupancy drops below the value in the initial state. Both observations suggest that a transient state is induced which is even more insulating than the initial state, as a result of reduced screening and an enhanced effective interaction. 

This is consistent with the dynamics of $\text{Re}[\mathcal{D}(\omega)]$  which is increased at low energies for at all times, see Fig.~\ref{Fig:MIT_bath}(e), so that electrons are moving in an effectively more insulating-like system. Similarly the spectral function at $\omega=0$ shows an initial decrease, which is followed by a slow re-filling of the pseudo-gap (see Fig.~\ref{Fig:MIT_bath}(c,g)).

The opposite procedure is to start in an initially metallic state and by applying an electric field pulse destroy the quasi-particle peak, 
which results in a reduced screening effect. 
Photo-doping an initially metallic state also leads to an increase in $\Delta E_\text{kin}$ (reduction in the absolute value of the kinetic energy), and after a brief transient a drop in the number of doubly occupied sites. The time resolved spectral function shown in panel (d) reveals a rapid destruction of the quasi-particle peak, which is followed on longer time-scales by a slow recovery. The destruction of the quasi-particle peak 
reduces the low-energy screening, 
see Fig.~\ref{Fig:MIT_bath}(f), since 
conducting electrons are removed from the system. Despite the slow reappearance of the quasi-particle peak, the values of $E_\text{kin}$ and $n_d$ at the longest times indicate that the system is considerably more strongly correlated than in the initial state, so that the system may again be regarded as trapped in a transient insulating nonequilibrium state. Despite the strong coupling to the heat-bath, the relaxation back to the inital metallic state appears to be inhibited on the accessible time-scales, presumably due to strong doublon-hole scattering. Note that this effect was already present in the system studied in the previous subsection, where the double occupancy $n_d$ for $U=7,\lambda=1$ was reduced below the initial value, see Fig.~\ref{Fig:nd_bath}(b), but there, due to the competing effect of the stronger photo-doping, the dynamics of screening was non-monotonous. In order to eliminate this competing effect we used here a weak pulse, which prevents a substantial photo-doping, but still results in the destruction of the quasi-particle peak.

\subsection{Reduction of the gap}
\label{Sec.:gap_reduction}

\begin{figure*}[ht]
\includegraphics{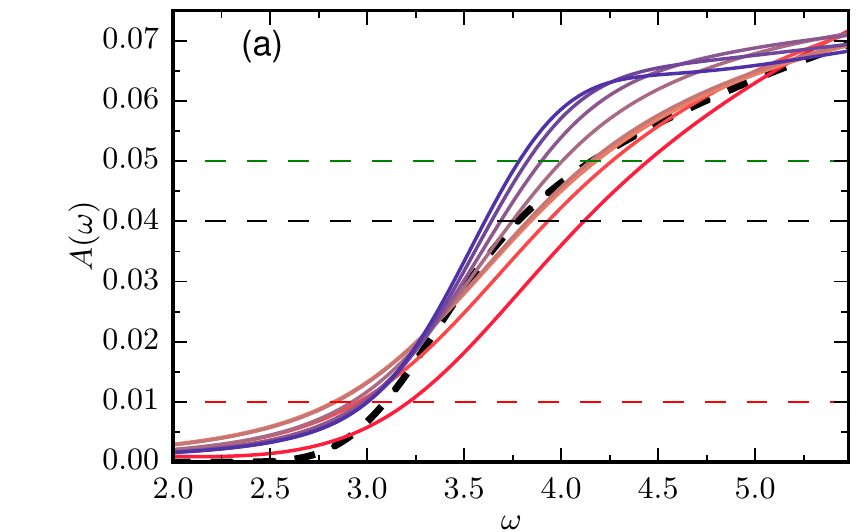} 
\includegraphics{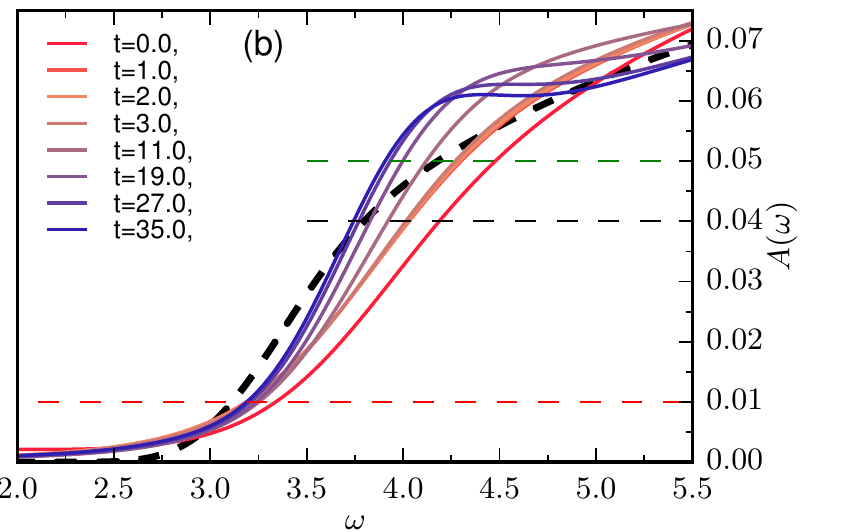} \\
\includegraphics{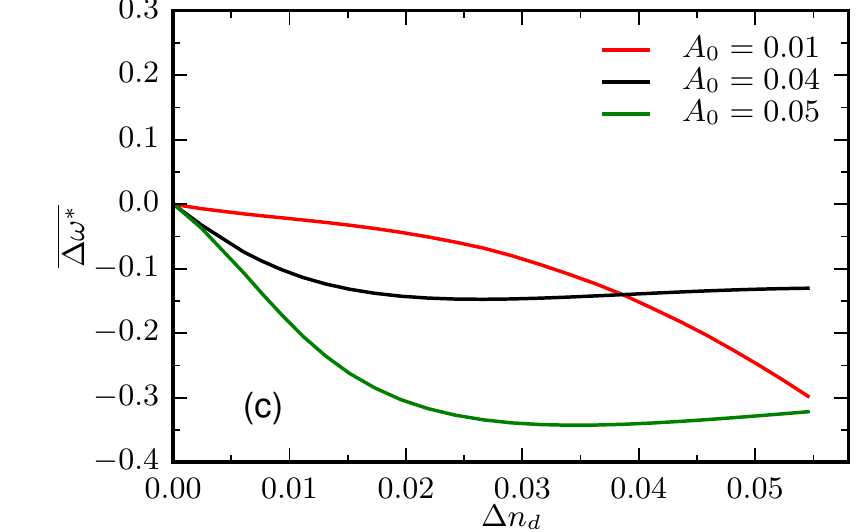}
\includegraphics{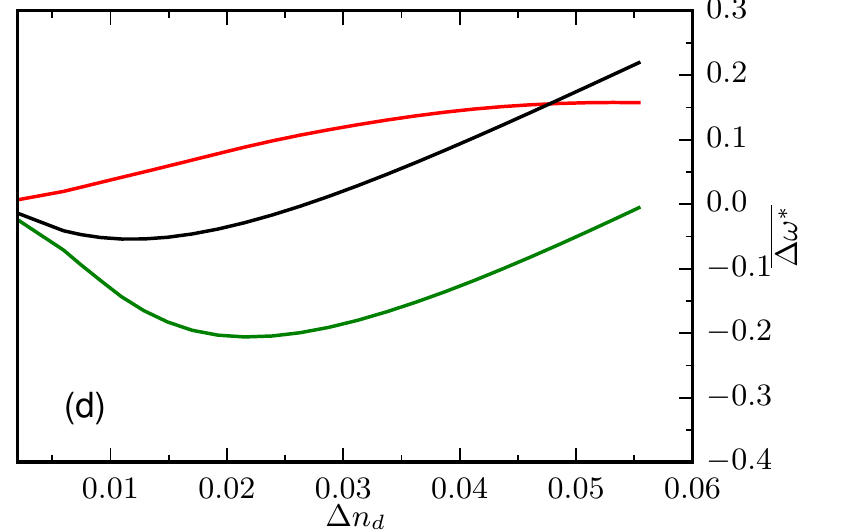}
\caption{ (a,b) Partial Fourier transform of the spectral function $A(\omega,t)$  for  $U = 14,V = 2,E_0 = 10.0,\omega_0=14$ (a) and $U = 14,V = 0,E_0 = 10.0,\omega=14$ (b) and coupling to a bath with $\lambda=1.0$. The dashed black lines represent the initial equilibrium spectral function. Change in the gap size $\overline{\omega^*}$ (see discussion in the main text) as a function of the change in the double occupancy $\Delta n_d $ for $V=2$ (c) and $V=0$ (d).
}
\label{Fig:Gap_reduction}
\end{figure*}

The effect of the screening on the reduction of the gap in the presence of the thermal bath was presented in Fig. \ref{Fig:spectral}(c)-(d) by the dashed lines, which indicate a gradual recovery of the gap back to its 
original size. In the case of $U=10$, see Fig.\ref{Fig:spectral}(c), and the cut at $A_0=0.01$ we see the decrease and subsequent reappearance of the gap, while the cut at higher energy ($A_0=0.03$) shows a decrease on the longest availble time scales. For the metallic solution $U=7$ the gap is also initially reduced, but for longer times it starts to recover. The different behavior of the two cases is the consequence of the longer thermalization times in the insulating case $U=10.$ 
In order to eliminate the effect of charge recombination we will study a system deep in the Mott insulator, namely $U=14.$ To present the effect of screening we will compare the evolution of the spectral function $A(t,\omega)$ for the Hubbard model ($V=0$) and the $U$-$V$ Hubbard model with $V=2$. The density of photo-doped carriers is $\Delta n_d\approx 0.04$. At the shortest times the charge carriers are inserted near the middle of the upper Hubbard band. While the pulse is still on, there is a strong distortion of the spectral function due to field induced effects.
The inclusion of the thermal bath enhance the subsequent relaxation of the high energy doublons to the lower edge of the Hubbard band. This leads to the formation of a pronounced shoulder-like feature at the band edge both for $V=0$ and $V=2$, in agreement with previous results for the Hubbard model.\cite{eckstein2013} In contrast to the Hubbard model, however, which shows only a formation of the shoulder-like feature, for $V=2$ the  increased screening effect leads to a reduction of the gap, see Fig. \ref{Fig:Gap_reduction}(a). 

Following the analysis in Sec.~\ref{Sec.:spectral} we analyze the frequency $\omega^*(t)$, where the spectral function takes some fixed value: $A(t,\omega^*(t))=A_0.$ The reduction of the gap $\overline{\Delta\omega^*}$ is defined as the difference between the long time average  $\overline{\omega^*}$ (measured between times $t=15$ and $t=25$) of the excited system and the equilibrium system $ \overline{\Delta \omega^*}=\overline{\omega^*}(E_0)-\overline{\omega^*_\text{eq}}$. The reduction of the gap $\overline{\Delta\omega^*}$ versus the change in the double occupancy $\Delta n_d$ for different pulse strengths between $4.0\leq E_0 \leq 15.0$ is shown in Fig.~\ref{Fig:Gap_reduction}(c). The model with screening $V=2$ shows a decrease of the gap for all cuts at different energies. In the case of low photo-doping $\Delta n_d\leq0.015$ the decrease of the gap is linear. The fast decrease and the later saturation for $A_0=0.04$, $0.05$ is a consequence of the formation of the shoulder like feature at the edge of the Hubbard band, while at lower energies $A=0.01$, $\overline{\omega^*}$ monotonously decreases with increasing density of the photo-doped carriers, due to the partial filling-in of the gap. Therefore in all of the analyzed cases the effect of screening is stronger than the distortion of the spectral function due to photo-doping. The Hubbard model shows a quite different behavior: the initial reshaping of the Hubbard band leads to an increase of the Hubbard gap for low energy cuts ($A_0=0.01$), while for $A_0=0.04,0.05$ the initial decrease in $\overline{\Delta\omega^*}$ is a consequence of the formation of the  shoulder like feature. The increase at even higher photo-doping is a consequence of the stronger reshaping of the upper Hubbard band.

\section{Conclusions}

We have discussed the nonequilibrium generalization of extended dynamical mean field theory, which allows to study the screening effect in correlated lattice systems in real time. To solve the nonequilibrium EDMFT equations, we have introduced a perturbative impurity solver which combines a self-consistenstent hybridization expansion with a weak-coupling expansion in the retarded interaction. This method should yield qualitatively correct results in Mott insulators or the metal-insulator crossover regime at elevated temperature, and for systems where the effect of screening is relatively weak (no tendency of charge-ordering). 

The formalism has been applied to the half-filled $U$-$V$ Hubbard model on the square lattice, which is driven out of a strongly correlated equilibrium state by a single-cycle laser pulse. The photo-doping of carriers into a Mott insulator leads to a rapid change in the screening environment, on the timescale of the inverse bandwidth, and an associated reduction in the size of the Mott gap. The low-energy bosonic excitations in the photo-doped Mott insulator also open up new relaxation channels, which can substantially reduce the doublon-hole recombination or production and hence the thermalization time of the system. 

In strongly correlated metallic systems, the effect of a field pulse depends on the pulse frequency and amplitude. In these systems, the low-energy quasi-particle band contributes to the screening. If the coherent quasi-particles are destroyed by a weak low-energy pulse, low-energy screening processes are eliminated, and the system can be switched into a more strongly interacting, hot insulating state. If the pulse intensity is large, or the pulse frequency is of the order of the splitting between the Hubbard side-bands, the resulting state is essentially a photo-doped narrow-gap insulator in which the screening from low-energy excitations within the photo-doped Hubbard bands may overcompensate the loss of screening associated with the vanishing of the quasi-particle peak. Even though the effective on-site interaction in such a photo-doped narrow gap insulator can be less than in the initial metallic equilibrium state, the re-emergence of a quasi-particle peak appears to be prevented by heating and strong doublon/hole scattering.  Within the timescale accessible in our simulations (about 30 inverse hoppings) we were only able to observe the emergence of a quasi-particle peak in systems coupled to a heat bath with broad energy spectrum, or in quenches to on-site interactions which are substantially lower than what can be achieved by photo-doping in the regime of weak-to-intermediate $V.$

For a more accurate simulation in the metal-insulator crossover regime and the study of screening induced transient states near the first order metal-insulator transition, a more accurate solver for the electron-boson impurity problem is needed. The extension of our NCA + weak coupling based solver to one crossing diagrams, or the implementation of the real-time OCA + Lang-Firsov scheme would seem the obvious next steps on the methodological side.

Our nonequilibrium EDMFT framework is an important step towards the development of an ab-initio scheme for out-of-equilibrium strongly correlated materials. The screening of the Coulomb interaction is a crucially important effect in solids, and simple schemes, such as a straight-forward nonequilibrium extension of the widely used LDA+DMFT framework, will fail to describe changes in the screening environment induced for example by a laser pulse. Ab-initio schemes such as the GW+DMFT method,\cite{biermann2003} which are built on top of an EDMFT framework and compute the (partially) screened Coulomb interaction in a self-consisten manner, appear to be a promising route forward.

\acknowledgements 

We thank L. Boehnke, A. Herrmann, L. Huang, Y. Murakami and H. Strand for helpful discussions.  
The calculations have been performed on the Beo04 cluster at the University of Fribourg. 
DG and PW acknowledge support from FP7 ERC starting grant No. 278023 and from SNSF Grant No. 200021\_140648.
\appendix

\section{Bosonic propagator from charge-charge correlations}\label{App.:Boson_from_chi}
In this appendix we derive the relation between the bosonic propagator $W$ and density-density correlator $\chi$ defined in Eq.~(\ref{Eq.:Bosonic_from_cc}) for the nonequilibrium case. 
Using the action (\ref{Eq.:Action_EDMFTwBoson}) the bosonic propagator $W$ can be expressed as~\cite{ayral2013} 
\begin{align}
W_\text{imp}(t,t')&=-2\frac{\delta \ln(Z)}{\delta \mathcal{U}^{-1}(t',t)} \nonumber\\
&=-2 \frac{\delta Z}{\delta \mathcal{U}(t_1,t_2)} * \frac{\delta \mathcal{U}(t_2,t_1)}{\delta \mathcal{U}^{-1}(t',t)} * \frac{1}{Z}\nonumber\\
&=2\mathcal{U}(t,t_1) * \frac{\delta \ln(Z)}{\delta \mathcal{U}(t_1,t_2)} * \mathcal{U}(t_2,t')
\end{align}
where we have used the chain rule and the relations $\mathcal{U}*\mathcal{U}^{-1}=\delta_\mathcal{C}$, $\delta \mathcal{U} * \mathcal{U}^{-1}=-\mathcal{U} * \delta \mathcal{U}^{-1}$, which imply 
\begin{align}
&\delta \mathcal{U} = - \mathcal{U} * \delta \mathcal{U}^{-1} * \mathcal{U}, \nonumber\\
& \frac{\delta \mathcal{U}(t_1,t_2)}{\delta \mathcal{U}^{-1}(t',t)} = -\mathcal{U}(t_1,t') \mathcal{U}(t,t_2).
\end{align}
On the other hand, we may can express the partition function using Eqs.~(\ref{Z_e}) and (\ref{Eq.:Action_EDMFT}). From this, we obtain  $\frac{\delta \ln[Z]}{\delta \mathcal{U}}=-\frac{1}{2}\chi_\text{imp}+\frac{1}{2}\mathcal{U}^{-1}$ and finally arrive at  
\beq{
  W_\text{imp}=\mathcal{U}-\mathcal{U}*\chi_\text{imp}*\mathcal{U}.
}

\bibliography{BibTex/tdmft,BibTex/Books,BibTex/Polarons}

\end{document}